\documentclass{aastex63}

\usepackage{ulem}
\usepackage{url}

\received{2021 January 22}
\revised{2021 April 8}
\accepted{2021 April 14}
\submitjournal{ApJ}

\shorttitle{MIR study of S208}
\shortauthors{Yasui et al.}

\begin{document}

\title{{\it Spitzer} Mid-infrared Study of Sh 2-208: Evolution of
Protoplanetary Disks in Low-metallicity Environments}

\correspondingauthor{Chikako Yasui}
\email{ck.yasui@gmail.com}

\author{Chikako Yasui}
\affil{National Astronomical Observatory of Japan, California Office, 
137 W. Walnut Ave., Monrovia, CA 91016, USA} 




\begin{abstract}

This study presents sensitive MIR photometry obtained with the {\it
Spitzer}/IRAC for a young cluster in Sh 2-208 (S208) located in one of
the lowest-metallicity \ion{H}{2} regions in the Galaxy, ${\rm [O/H]} =
-0.8$ dex.
 Previous studies suggested that the cluster is $\sim$0.5-Myr old and
 has a distance of $D = 4$ kpc, which is consistent with the astrometric
 distance from Gaia EDR3. In $\sim$$3.5 \times 4$-arcmin field, 96
 sources were detected in at least one MIR band at $\ge$10$\sigma$,
 covering intermediate-mass stars with $\sim$1.0-$M_\odot$
 {mass detection} limit.
Total 41 probable cluster members were identified based on the
spatial distributions of spectral-energy-distribution slopes derived
from the NIR $K_S$ and IRAC bands and extinctions of the sources.
The cumulative distribution of the SED slopes for the S208 cluster was
not significantly different from those of other clusters in
solar-metallicity environments with approximately the same age for
intermediate-mass stars, if one also considers
{non-detected} MIR sources identified as S208 cluster
members from NIR observations. This suggests that the degree of dust
growth/settling does not significantly change with metallicities as
different as $\sim$1 dex.
The fraction of stars with MIR disk emissions for the cluster members
with $\ge$1-$M_\odot$ mass was 64\%--93\%, which is comparable to the
results in solar-metallicity environments. Although this may suggest
that dominant disk-dispersal mechanisms for intermediate-mass stars have
either no, or very weak, dependence on metallicity, it can be argued
alternatively that this may suggest that the {disk
dispersal} process does not work effectively at this young stage.

\end{abstract}





\section{Introduction} \label{sec:intro}

Protoplanetary disks are the sites of planet formation; therefore, their
evolution is critical for understanding planet formation.
The presence of disks was first confirmed in silhouettes from optical
observations obtained with the Hubble Space Telescope {\it Hubble Space
Telescope} ({\it \hspace{-0.5em} HST}) \citep[e.g.,][]{O'dell1993}.
%
Beginning in the 2000s, disks were observed as color excesses
at near-infrared (NIR) wavelengths, and disk lifetimes were derived for
the first time to be as short as $\sim$5--10 Myr \citep{Lada1999}.
After the advent of the {\it Spitzer Space Telescope}, disk lifetimes
were estimated to be almost the same from {mid-infrared
(MIR)} observations, but the observations also made it possible to
determine the evolutionary stages of the disks \citep{Lada2006}.
Recent ALMA (Atacama Large Millimeter Array) observations enabled the
disk investigations of various star-forming regions, making it possible
to detect the entire disk with {very high} spatial resolution.
However, because great diversity was found in disks, e.g., in disk
masses \citep{Ansdell2017} and disk structures \citep{Andrews2018}, 
a comprehensive understanding of protoplanetary disks, and thus planet
formation, seems to be increasingly difficult.

Instead of observing nearby protoplanetary disks in more detail, we have
been studying disk evolution focusing on specific physical parameters
such as stellar mass \citep{Yasui2014} and metallicity
\citep{Yasui2010}.
These two parameters are thought to greatly affect the
probability of planet formation (e.g., for giant planet formation;
\citealt{Johnson2010}).
The present study focuses on metallicity.
Defined as the proportion of material comprising elements other than
hydrogen or helium, metallicity is known to increase with cosmic
evolution due to the synthesis of elements in stars and
supernovae.
Presently, only 2\% of the mass of baryons in our solar
system comprise these heavy elements.
Nevertheless, metallicity is
believed to be one of the most critical factors for planet formation
because dust grows into planetesimals, which are the building blocks of
planets, despite being present in the disk at only 1\% by mass.
Metals
also very sensitively affect the heating and cooling accompanying
planet-forming processes, and they directly affect radiative
transfer.
{Whether or not any dependence of disk evolution on
metallicity is found from observations can have great influence on the
theory of planet formation}
(e.g., \citealt{Yasui2009}, \citealt{Ercolano2010}).

Some observations were previously performed for regions with other than
solar metallicity.
\citet{Maercker2005} presented ground-based presented a ground-based
 $L$-band (3.5 $\mu$m) imaging of 30 Doradus in the Large Magellanic
 Cloud (LMC).
From the {\it Spitzer} SAGE (Surveying the
Agents of a Galaxy's Evolution) survey of the LMC and Small Magellanic
Cloud (SMC) \citep{{Meixner2006},{Gordon2011}}, a very large number of
new young stellar object (YSO) candidates were identified in the LMC and
SMC.
\citet{De Marchi2010} and {their group} performed {\it
HST} optical imaging, including an H$\alpha$ band, of six star-forming
regions in the LMC and SMC.
However, the {mass detection} limit of these studies was
relatively high due to the large distances {of}
the LMC and SMC ($D \sim 50$ kpc).
Other factors besides metallicity may also influence the evolution of
protoplanetary disks in the case of {extragalactic
target} observations.

Compared to external galaxies, Galactic targets have great advantages
for detecting lower-mass sources, due to the smaller distances involved
($D < 10$ kpc).
We focus here on a wide range of metallicities, extending from
$\sim$$-$1 to $\sim$$+$0.5 dex, within the Galaxy.
In a series of papers, we have compiled a list of low-metallicity (${\rm
[O/H]} \simeq -1$ dex) young clusters in the Galaxy and have presented
deep NIR imaging observations with a {mass detection}
limit of $\sim$0.1 $M_\odot$.
The estimated fractions of stars with disks based on NIR continuum
emission (NIR disks) in such clusters are lower than those in the solar
neighborhood, suggesting that the lifetimes of protoplanetary disks in
such environments are shorter than in clusters with solar metallicity
\citep{{Yasui2009}, {Yasui2010}, {Yasui2016_S207}, {Yasui2016_S208},
{Yasui2021}}.

\citet{Puga2009}, \citet{Cusano2011}, and \citet{Kalari2015} studied
another \ion{H}{2} region in the Galaxy, Sh 2-284
{(S284)}, which is located at a distance $D \sim 4$ kpc,
has a high-mass star-forming region, and is possibly in a
low-metallicity environment.
\citet{Cusano2011} presented optical spectroscopic data, including the
H$\alpha$ band, and photometric data from the optical band to the {\it
Spitzer}/IRAC MIR band, and they estimated the age of a star-forming
cluster (Dolidze 25) located in the center of S284 to be 2 Myr.
They suggested that a significant fraction of stars in the cluster still
has disks that are traced with H$\alpha$ emissions and with MIR
continuum emissions (MIR disks) with a {mass detection}
limit of $\sim$1 $M_\odot$.
\citet{Kalari2015} {suggested} no evidence for a
systematic change in the mass accretion rate with metallicity from
optical spectroscopy.
However, note that {S284} is a very large extended region
spanning $\sim$20 pc, and sequential star-forming activities are
suggested here \citep{Puga2009}.
In the solar neighborhood, the sizes of young clusters are typically $\sim$1 
pc \citep{Adams2006}.
For a direct comparison of results in low-metallicity environments with
those in the solar neighborhood, observations of individual clusters on
similar spatial scales are essential.
Moreover, note that different wavelengths trace the different parts of
protoplanetary disks: NIR and MIR continuum emissions trace heated dust
located in the inner parts of disks, whereas H$\alpha$ emissions trace gas
accreting from disks onto the central stars.

For the next step, we aim to perform MIR studies of individual
star-forming cluster in low-metallicity environments.
This will enable us to investigate the metallicity dependence of disk
evolution by comparing the results with those obtained for clusters in
the solar neighborhood.
Because the MIR disk traces a slightly larger radius in a protoplanetary
disk (a stellocentric distance of approximately a few AU), compared to
the NIR disk, which traces the inner disk edge (a stellocentric distance
of $\sim$0.1 AU), it may be possible to investigate the properties of
disks more precisely in low-metallicity environments.
From our compilation of star-forming clusters in low-metallicity
environments, all of which have cluster radii of $\sim$1 pc (see Figure~%
12 in \citealt{Yasui2021}), Sh 2-208 (S208) is found to have a
relatively small distance ($D\sim 4$ kpc).
Among such regions, S208 turned out to be almost the only cluster with
sensitive MIR imaging obtained with {\it Spitzer}/IRAC.
S208 \citep{Sharpless1959} is one of the lowest-metallicity \ion{H}{2}
regions in the Galaxy, with ${\rm [O/H]} = -0.8$ dex
\citep[e.g.,][]{Rudolph2006}.
\citet{Bica2003} identified a star-forming cluster in this \ion{H}{2}
region, whereas \citet{Yasui2016_S208} identified 89 members to a
detection limit of $\sim$0.2 $M_\odot$ from deep NIR images obtained
with Subaru/MOIRCS (Multi-Object InfraRed Camera and Spectrograph).
{From the fitting of the {\it K}-band luminosity function
(KLF)},
the cluster age and distance were estimated to be $\sim$0.5 Myr and
$\sim$4 kpc, respectively.
Despite the very young age, the disk fraction of the cluster was
estimated to be only $\sim$30\%, which is significantly lower than those
in solar-metallicity regions of approximately the same age.
These investigations also suggested from large-scale MIR images that
sequential star formation is occurring in the region surrounding S208,
triggered by an expanding bubble with a $\sim$30-pc radius.

The remainder of this study is organized as follows: Section 2 discusses
the distance {of} S208; Section 3 explains the
observations and data reduction; Section 4 presents the
{obtained images and detected sources}, derives their
spectral energy distributions (SEDs), and identifies the cluster
members; and Section 5 classifies the SED slopes of the cluster members
into evolutionary stages of the protoplanetary disks, derives the MIR
disk fraction of the cluster members, compares the results of S208 to
those of the clusters in the solar neighborhood and in other
low-metallicity environments, and finally discusses the implication of
protoplanetary disk evolution.


\section{Distance of S208} \label{sec:distance}

The distance {of} S208 was very recently updated from the
Gaia Early Data Release 3 (Gaia EDR3; \citealt{Gaia2020}), and previous
distance estimates for S208 were summarized in Section~2.1 of
\citet{Yasui2016_S208}.
The parallax of the probable dominant exciting source of S208, GSC
03719-00517, was measured with Gaia EDR3, and the obtained astrometric
distance is $D=4.8^{+0.5}_{-0.4}$ from the parallax $p = 0.21 \pm 0.02$
mas (Gaia source ID 275284696186500096).
S208 is suggested to be located in a {sequential
star-forming} region (see Figure~13 in \citealt{Yasui2016_S208}).
Sh 2-207 (S207) is another \ion{H}{2} region located in this
{sequential star-forming} region, very close to S208, at
a separation of $\sim$10 arcmin, and the astrometric distance
{of} the probable dominant exciting star of S207, GSC
03719-00546 \citep{Yasui2016_S207}, was measured from the Gaia ERR3 to
be $D=3.8 \pm 0.3$ kpc from the parallax $p=0.26\pm0.02$ mas (Gaia
Source ID 275291258896477312).
There is another open cluster in this {sequential
star-forming} region, the Waterloo 1 cluster (labeled ``Wat1'' in
Figure~13 of \citealt{Yasui2016_S208}).
\citet{Cantat-Gaudin2008}, using Gaia Data Release 2 (DR2), established
the most likely distance {of this cluster} to be 4.1 kpc.
Considering the {sequential star-forming} activities
suggested by \citet{Yasui2016_S208}, it is unlikely that only S208 is
located in a different region than S207 and Waterloo 1.
\citet{Vallee2020} also suggested that all these regions are located in
{an interarm} island between the Perseus arm and the
Cygnus arm.
The stellar density {of S208} is very
 high, and the source may not be sufficiently resolved. In fact, the
 accuracy of the observations had been low until DR2, which yielded
 $D=6.25^{+1.44}_{-0.99}$ kpc from the parallax of $0.16 \pm 0.03$ mas,
 compared to the other open clusters (S207 and Wat 1). This may cause
 some offset in the distance estimate of S208.

From the fitting of the KLF, \citet{Yasui2016_S208} suggested that the
kinematic distance ($D=4$ kpc) is more likely than the photometric
distance ($D=8$--9 kpc) for both S208 and S207 (see
\citealt{Yasui2016_S207}), and they estimated the age of S208 to be
$\sim$0.5 Myr.
The kinematic distance was estimated to be $\simeq$4 kpc, with a radial
velocity of $\simeq$$-$30 km s$^{-1}$, from observations of H$\alpha$,
\ion{H}{1}, and $^{12}$CO (references in \citealt{Yasui2016_S207} and
\citealt{Yasui2016_S208}) for both S207 and S208.
In the direction of S208 ($l = 151.3^\circ$, $b = +2.0^\circ$), the
difference in the astrometric distances from Gaia EDR3 between S208 and
S207 ($\sim$1 kpc) corresponds to a difference of $\sim$7 km s$^{-1}$ in
radial velocity, based on the method of \citet{Reid2009}.
{This} is a much larger velocity difference than the
actual differences from the observations.
This supports the conclusion that S208 is also located in the same
{sequential star-forming} region ($D \simeq 4$ kpc) as
S207 and Waterloo 1.

Therefore, we have adopted herein 4 kpc as the distance
{of} S208.
There is no significant difference in the age of S208, which was a
parameter estimated at the same time as the distance by
\cite{Yasui2016_S208}, even if S208 exists in a background region
different from the {sequential star-forming} region if
the distance estimate {of} S208 from Gaia EDR3 is correct
($D=4.8$ kpc) because the difference in the distance modulus is very
small, $<$0.5 mag, and the model KLFs used in the KLF fitting do not
change significantly.
We will discuss the impact of these assumptions in the subsequent
sections.
By focusing on intermediate-mass stars, the discussion in this paper is
not significantly affected by the difference in the distance between 4
kpc and 4.8 kpc.
In the near future, whether or not all of these clusters exist in the
same {sequential star-forming} region will be clarified                                                                           
by considering the more precise data that will be released as the
astrometric survey by Gaia progresses.


\section{Observations and Data Reduction} 

\subsection{{\it Spitzer}/IRAC Mid-IR Archival Data and Photometry}\label{sec:Spitzer}

The {\it Spitzer Space Telescope} is an {85 cm-diameter}
telescope with an infrared array camera (IRAC) covering bands centered
at 3.6, 4.5, 5.8, and 8.0 $\mu$m \citep{Fazio2004}.
We obtained {four band IRAC} images of S208 from archival
data obtained on October 16, 2007 (UT), for which the program ID is
30734.
A five-position Gaussian dithering pattern was used, and the resulting
images in each wavelength band covered a $5\arcmin \times 5\arcmin$
area.
The total integration time was 471 s in each of the four bands
\citep{Richards2012}.
We used the post-basic calibrated data from the Spitzer Heritage Archive
(SHA)\footnote{\url
http://sha.ipac.caltech.edu/applications/Spitzer/SHA/}, which were
reduced with the data-reduction pipeline.
The images had a full width at the half maximum ${\rm (FWHM)} =
1.6$--1.9 arcsec from 3.6 to 8.0 $\mu$m, with a pixel scale of
0.6$\arcsec$ pixel$^{-1}$.
We performed photometry for the {four band IRAC images}
in the same field as the S208 region {centered} at
$\alpha_{2000} = 04^{\rm h} 19^{\rm m} 36^{\rm s}$, $\delta_{2000} =
+52^\circ 58\arcmin 58\arcsec$, with dimension of $\sim$$3.5\arcmin
\times 4\arcmin$.
Observations with high spatial resolution and sensitivity were obtained 
(Section~\ref{sec:moircs}).

\begin{figure}[!h]
\begin{center}
\includegraphics[scale=1]{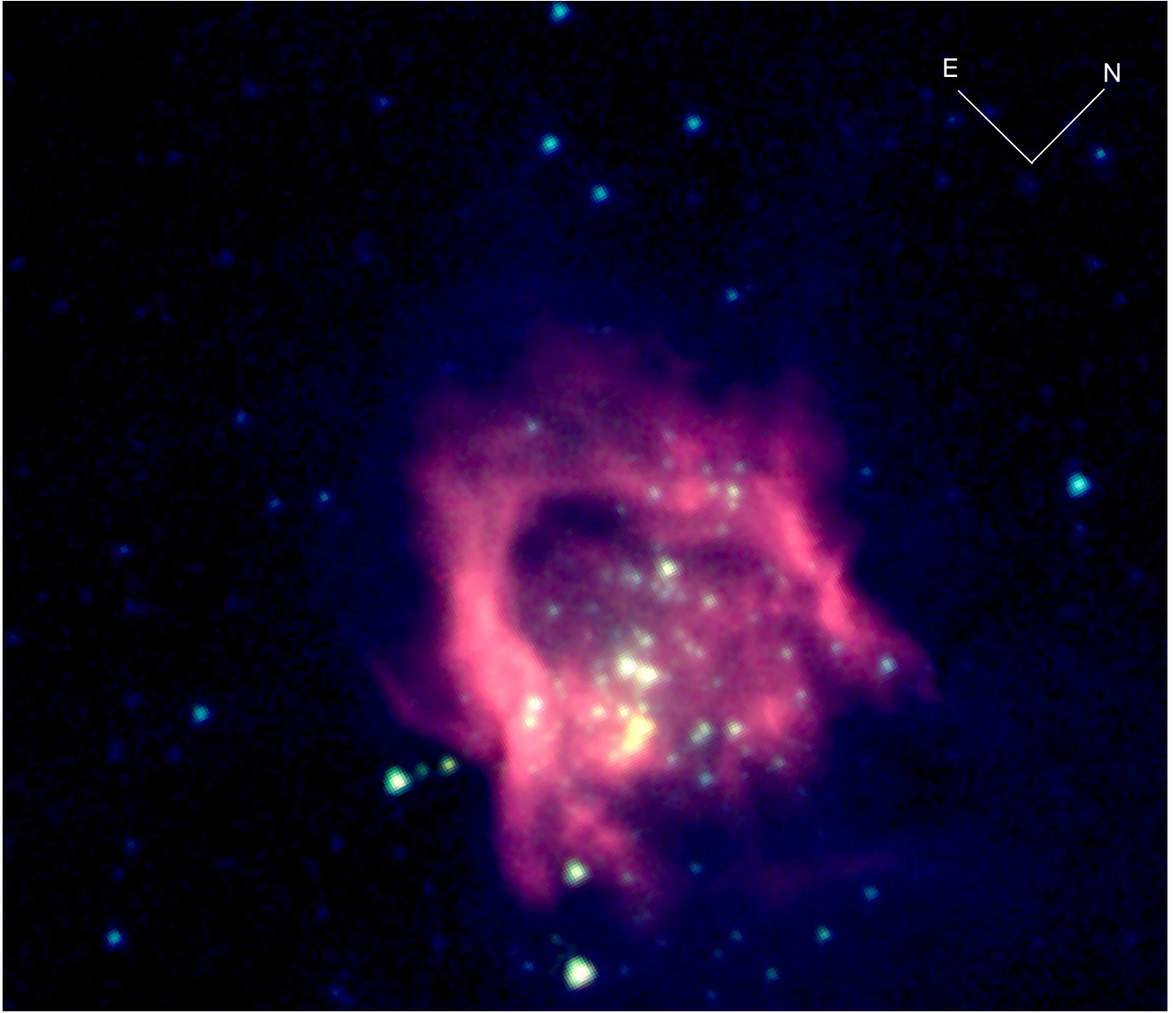}
\caption{Pseudocolor image of S208, with the field centered at
$\alpha_{2000} = 04^{\rm h} 19^{\rm m} 36^{\rm s}$, $\delta_{2000} =
+52^\circ 58\arcmin 58\arcsec$ in equatorial coordinates.
The color image was produced by combining the {\it Spitzer}/IRAC 3.6
$\mu$m (blue), 4.5 $\mu$m (green), and 5.8 $\mu$m (red) images.
The field view of the image is $\sim$$3.5\arcmin \times 4\arcmin$.
The {Galactic} longitude is along the x-axis, and the
 {Galactic} latitude is along the y-axis.}
\label{fig:S208cl_Spitzer}
\end{center}
\end{figure}

Figure~\ref{fig:S208cl_Spitzer} shows a pseudocolor image of the
{\it Spitzer}/IRAC bands.
The target cluster is very crowded, and many stars are detected in the
3.6- and 4.5-$\mu$m band images; thus, we performed photometry for
{two band} data using point-spread function (PSF) fitting
obtained with IRAF/DAOPHOT\footnote{IRAF is distributed by the National
Optical Astronomy Observatories, which are operated by the Association
of Universities for Research in Astronomy, Inc. under a cooperative
agreement with the National Science Foundation.}.
To derive the PSF, we selected bright stars separated by $r>1$ arcmin,
but within 3 arcmin from the cluster center derived in
\citet{Yasui2016_S208},
$\alpha_{2000} = 04^{\rm h}19^{\rm m}32.7^{\rm s}$, $\delta_{2000} =
+52^\circ 58\arcmin 34.6\arcsec$.
The PSF photometry was performed twice using the ALLSTAR routine, i.e.,
once using the original images and a second time using the images with
sources obtained from the first iteration subtracted.
We used PSF fitting radii of 3 pixels for both the 3.6- and 4.5-$\mu$m
band images, which are the PSF FWHM values, and we used sky annuli with
inner radii and widths respectively four and three times as large as the
PSF fit radii.
Finally, aperture correction was performed using isolated bright stars
detected with more than 10$\sigma$ and brighter than 14.5 mag in both
the 3.6- and 4.5-$\mu$m band images.
For the 5.8- and 8.0-$\mu$m band images, the number of sources seen on
the images was very small ($<$10).
Aperture photometry was then performed. We adopted zero-point magnitudes
and the standard aperture correction (radius on the source of
2.4$\arcsec$ and background annulus of 2.4$\arcsec$--7.2$\arcsec$) from
the IRAC Instrument
Handbook\footnote{\url{http://irsa.ipac.caltech.edu/data/SPITZER/docs/irac/iracinstrumenthandbook/}}
for all IRAC bands.
We selected these relatively small apertures and sky annuli to obtain
the best possible subtraction of background emission from the
nebulosities in the star-forming regions, which are bright and spatially
variable at MIR wavelengths \citep{Lada2006}.
The limiting magnitudes (10$\sigma$) based on the pixel-to-pixel noise
for the 3.6 and 4.5 $\mu$m-band data were estimated to be $\simeq$15.5
and $\simeq$15.0 mag, respectively. Limiting magnitudes cannot be
estimated for the 5.8- and 8.0-$\mu$m bands data because the numbers of
detected sources were very small ($<$10).

\subsection{Near-IR Photometry from the Literature} \label{sec:moircs}

In addition to the MIR data, NIR {\it JHK}-band data are necessary in
this study {to estimate the extinction and mass for each
source and to determine the SED slope for each source for categorization
into evolutionary disk stages.} 
The stellar density in S208 is relatively high
(\citealt{Yasui2016_S208}; see also Figure~12 of \citealt{Yasui2021});
therefore, the archival data do not provide sufficiently accurate
photometry (e.g., 2MASS) due to inadequate spatial resolution (FWHM of
$\sim$2.5 arcsec).
We used the results of NIR {\it JHK}-band photometry from images with
spatial resolutions $\lesssim$1 arcsec in the literature
\citep{Yasui2016_S208}.
The investigators obtained deep imaging with the 8.2 m Subaru telescope
equipped with MOIRCS (\citealt{Ichikawa2006},
\citealt{Suzuki2008})\footnote{The images are available from the
Subaru-Mitaka-Okayama-Kiso Archive System (SMOKA): {\url
https://smoka.nao.ac.jp}.}.
MOIRCS has two 2K ``HAWAII-2'' imaging arrays, each chip covering $3.5'
\times 4'$, and it has a $4' \times 7'$ field of view.
However, because an engineering-grade detector was used during the
observation period, { which was exchanged with one of the
MOIRCS science-grade detectors,} \citep{Yasui2016_S208} presented only
the data obtained with another detector covering $3.5' \times 4'$, which
are shown in the northeastern half of their Figure~3.
{The center of the images for the two detectors was set
at $\alpha_{2000} = 04^{\rm h} 19^{\rm m} 45^{\rm s}$, $\delta_{2000} =
+53^\circ 05\arcmin 41\arcsec$, which corresponds to the center of the
science-grade detector of $\alpha_{2000} = 04^{\rm h} 19^{\rm m} 36^{\rm
s}$ $\delta_{2000} = +52^\circ 58\arcmin 58\arcsec$ covering the whole
S208 \ion{H}{2} region.} 
%
Aperture photometry, with an aperture diameter of 0.7$\arcsec$, was
performed for all the {\it JHK}-band images with a PSF FWHM
$\lesssim$1$\arcsec$ to avoid contamination by adjacent stars.
Consequently, 575 point sources were detected, and the limiting
magnitudes (10$\sigma$) were $J = 19.8$ mag, $H = 18.5$ mag, and $K_S =
18.0$ mag (Section~3 of \citealt{Yasui2016_S208}).


\section{Results} \label{sec:results_s208}

\subsection{Detection of Point Sources in IRAC Images} \label{sec:detect}

We detected 96 point sources at more than 10$\sigma$ in at least one of
the four IRAC bands.
A total of 79, 74, nine, and two sources were detected in the 3.6-,
4.5-, 5.8-, and 8.0-$\mu$m bands, respectively.
Figure~\ref{fig:disk_dist} depicts the spatial distribution of the
detected sources (squares), and Table~\ref{tab:S208cl_Spi} shows the
results of IRAC photometry along with NIR photometry from
\citet{Yasui2016_S208}.
Among the 96 sources detected in the IRAC bands, 95
{sources} were detected in at least one of the NIR {\it JHK} 
bands (Section~\ref{sec:moircs}), and only one source was not detected
in any of the NIR bands (ID 59 in Table~\ref{tab:S208cl_Spi}).

To estimate the {mass detection} limit of the IRAC
photometry and to select stars in a given mass range in later
sections{, we} plotted the sources detected in the IRAC
and NIR bands on the NIR $J-K_S$ vs. $K_S$ color-magnitude diagram
(Figure~\ref{fig:CM_S208}).
The sources detected in both bands are depicted by squares, whereas
those detected only in the NIR bands are shown {with}
dots.
The locus of dwarf main-sequence stars with solar metallicity in
spectral types O9 to M6 (corresponding to the masses $\sim$20--0.1
$M_\odot$) from \citet{Bessell1988} is shown as a black line.
The isochrone model for the age of the S208 cluster (0.5 Myr old) with
extinction of $A_V = 0$ mag is denoted by a thick blue line.
The isochrone models are from \citet{Lejeune2001} for the masses
$M/M_\odot \ge 7$, from \citet{Siess2000} for the masses $3 \le
M/M_\odot \le 7$, and from \citet{{D'Antona1997},{D'Antona1998}} for the
masses $M/M_\odot \le 3$.
The tick marks on the thick blue line show the position of the isochrone
models for the masses 20, 10, 5, 3, 1.5, 1, 0.5, and 0.1 $M_\odot$, with
the blue numbers showing the masses.
A 4 kpc distance was assumed.
The short arrow shows the reddening vector {of} $A_V = 2$
mag from the isochrone model {for} the mass of 20
$M_\odot$.
The black dashed lines are from isochrone tracks for the mass of 1 and
0.5 $M_\odot$ and parallel to the reddening vector.
Stars plotted above those lines {are indicated to have
masses larger than} the mass of the respective line.

Figure~\ref{fig:CM_S208} shows that a large fraction of sources with
masses $\ge$1 $M_\odot$ were detected in the IRAC bands, and a large
fraction of the sources detected in the IRAC bands had masses $\ge$0.5
$M_\odot$.
Approximately {one third} of the sources with suggested
masses $\ge$1 $M_\odot$ were not detected.
We checked them in the IRAC bands and found that the
{non-detection} was mainly due to insufficient spatial
resolution of the MIR {\it Spitzer}/IRAC images compared to the NIR
Subaru/MOIRCS images.
On the higher-mass side, no sources were bright enough to cause
saturation in the NIR bands, but they were detected in the IRAC bands.
Therefore, the mass range of the sources detected in this study was
$\sim$1--10 $M_\odot$, which covers the entire intermediate-mass range
{and slightly reaches low-mass stars.}

Note that we assumed $D=4.0$ kpc as the distance {of}
S208 although there is another possibility{, $D=4.8$
kpc,} which was obtained from the recent Gaia Data Release (EDR3)
(Section~\ref{sec:distance}).
There is a difference of 0.4 mag in the distance modulus between $D=4.0$ 
and 4.8 kpc.
In Figure~\ref{fig:CM_S208}, the gray dashed line shows the result for
1.5 $M_\odot$ assuming $D=4.8$ kpc; it is located at approximately the
same location as the black dashed line for 1.0 $M_\odot$ in the case
{of} $D=4.0$ kpc, which is the {mass
detection} limit in this study, albeit being located slightly above
{the black dashed line.}
Therefore, the difference in distances does not affect the following
discussions, which concentrate mainly on intermediate-mass stars.

\subsection{Spectral Energy Distributions} \label{sec:sed}

We derived the SED slope [$\alpha = d \log (\lambda F_\lambda) / d \log
(\lambda)$] for each source identified in Section~\ref{sec:detect} from
the NIR $K_S$ and IRAC MIR bands detected at more than 10$\sigma$.
MOIRCS, with which the NIR images were obtained in the literature, uses
the Mauna Kea Observatories (MKO) filter system \citep{Tokunaga2002};
therefore, we used the passband and {flux density}
conversions for the NIR data for the system from \citet{Tokunaga2005}.
We used the values by \citet{Reach2005} for the IRAC bands.
We determined two kinds of SED slopes: 1) SED slopes derived directly
from the observed data for each source ($\alpha$); and 2) SED slopes
derived from the data corrected for extinction using the $A_V$ derived
for each object ($\alpha_0$).
The extinction of each source was derived using the $J-H$ vs. $H-K_S$
color-color diagram (Figure~\ref{fig:CC_S208}) following the same
procedure as in \citet{Muench2002} {by dereddening} along
the reddening vector to the {young star} locus in the diagram.
Figure~\ref{fig:CC_S208} plots the sources detected in the IRAC MIR
bands as squares in the NIR color-color diagram.
For convenience, the {young star} locus was approximated
{by the extension of the classical T Tauri star (CTTS)
locus, as shown with} the gray line.
Only stars above the CTTS locus were {used.} 
The $A_V$ values were derived based on the distance required for
dereddening.
We defined stars below the locus as those with $A_V = 0$ mag.
The CTTS locus originally derived by \citet{Meyer1997} in the CIT system
and converted to the MKO system \citep{Yasui2008} was used.
The reddening law from \citet{Rieke1985} was employed.
To derive $\alpha_0$, we corrected the NIR and MIR magnitudes using the
reddening laws from \citet{Rieke1985} and from \citet{Indebetouw2005},
respectively, using the estimated $A_V$ values.
Table~\ref{tab:S208cl_Spi} lists the derived values of $\alpha$,
$\alpha_0$, and $A_V$.
The method of deriving $A_V$ values can be applied only to sources for
which photometric data are available in all the $J$, $H$, and $K_S$
bands; therefore, only the $\alpha$ value is shown for sources for which
this condition is not met (ID 44 and 59).

\subsection{Spatial Distribution and Membership Selection}
\label{sec:member}

\begin{figure*}[!h]
\begin{center}
\includegraphics[scale=1]{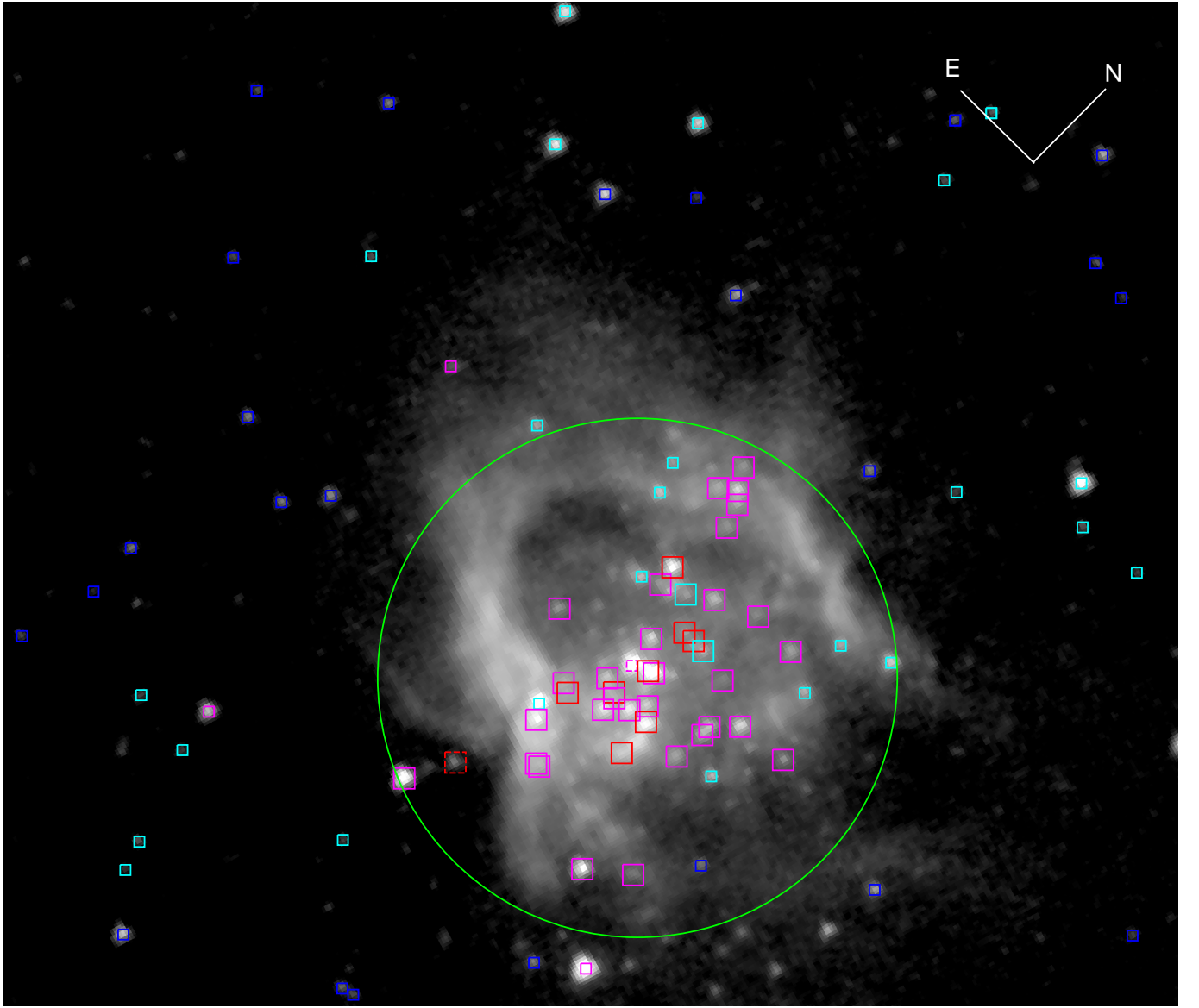}
\caption{Spatial distribution of the sources detected in the IRAC bands,
shown {with} squares superposed on an IRAC 3.5-$\mu$m image with the
same field of view as Figure~\ref{fig:S208cl_Spitzer}.
The different colors of the squares show different disk evolutionary
stages based on the SED slopes: red, magenta, cyan, and blue show class
I objects, class II objects, objects with evolved disks, and class III
objects, respectively.
Thick squares show sources for which $\alpha_0$ can be derived, and the
colors represent the $\alpha_0$ values.
The dashed squares show sources for which $\alpha_0$ cannot be derived,
and the colors represent the $\alpha_0$ values.
The large squares show probable cluster members identified in this
study, whereas the small squares show possible non-members.
The green circle outlines the cluster location, which has the central
coordinates
$\alpha_{2000} = 04^{\rm h}19^{\rm m}32.7^{\rm s}$, $\delta_{2000} =
+52^\circ 58' 40.5\arcsec$, 
and a radius of 50$\arcsec$.}
\label{fig:disk_dist}
\end{center} 
\end{figure*}

We categorized each source identified in Section~\ref{sec:detect}
according to the $\alpha_0$ values derived in Section~\ref{sec:sed}
based on the evolutionary stages of the protoplanetary disks.
\citet{Lada2006} and \citet{Hernandez2007} defined four categories using
the SED slopes: 1) $\alpha_0 > 0$ for class I candidates; 2) $-1.8 <
\alpha_0 \le 0$ for class II stars; 3) $-2.56 \lesssim \alpha_0 < -1.8$
for stars with evolved disks; and 4) $\alpha_0 < -2.56$ for class III
stars.
Accordingly, $\alpha$ values were used for those sources for which
$\alpha_0$ values cannot be derived (ID 44 and 59).
Figure~\ref{fig:disk_dist} shows the spatial distribution of the sources
detected in the IRAC bands, with different colors representing the
different categories.
The red, magenta, cyan, and blue squares show class I and II objects,
objects with evolved disks, and class III objects, respectively.
Figure~\ref{fig:disk_dist} shows that class I and II objects are
concentrated in the region surrounded by {a} green circle
centered at $\alpha_{2000} = 04^{\rm h}19^{\rm m}32.7^{\rm s}$,
$\delta_{2000} = +52^\circ 58' 40.5\arcsec$ and with a radius of
50$\arcsec$.
This suggests that the cluster extends to this field.
From the sources detected in the NIR bands by \citet{Yasui2016_S208}, we
also confirmed that no significant difference was present in the spatial
distribution of stellar density depending on direction.

\citet{Yasui2016_S208} estimated the coordinate of the cluster center
and the extent of the cluster from the projected stellar density using
the point sources detected in the NIR {\it JHK} bands.
From the densest point, they estimated the coordinates of the cluster
center to be $\alpha_{2000} = 04^{\rm h}19^{\rm m}32.7^{\rm s}$,
$\delta_{2000} = +52^\circ 58' 34.6\arcsec$. 
Although this center coordinate is 5.9$\arcsec$ off from the one
obtained herein, the difference is almost within the errors, which were
estimated in \citet{Yasui2016_S208}, $\sim$5$\arcsec$.
In their Figure~4, the stellar density decreased as the distance from
the center increased up to 64$\arcsec$ (corresponding to 550 pixels in
their images), and the density increased and decreased at greater
distances.
They rigorously estimated the extent of the cluster region to be
34$\arcsec$ (corresponding 300 pixels in their images) in radius, where
the stellar density was more than 3$\sigma$ above the entire sky frame.
However, because the green circle is located almost inside a circle with
a radius of 55$\arcsec$ around the cluster center determined by
\citet{Yasui2016_S208}, the stellar density in this region is also
considered to be more than that of the background level in their figure.
Therefore, we defined the cluster radius to be 50$\arcsec$ herein, which
corresponds to 1.0 pc assuming $D=4.0$ kpc to be the distance
{of} S208.
A total of 51 sources was detected in at least one of the IRAC bands
located in the cluster region.

\begin{figure}[!h]
\begin{center}
\includegraphics[scale=0.55]{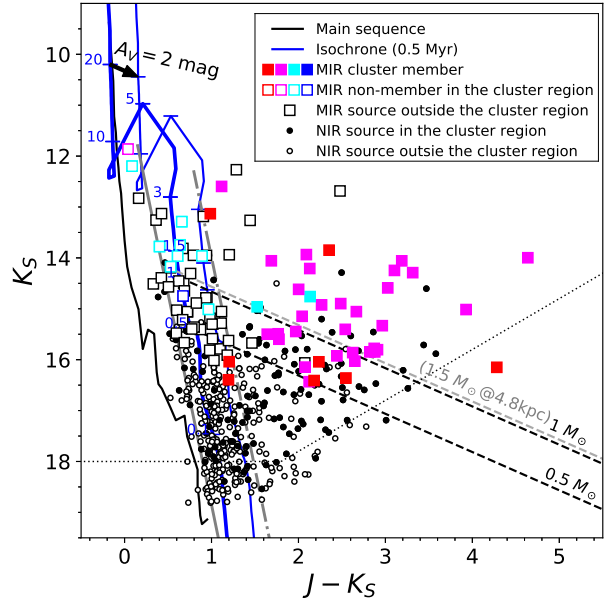}
\caption{ $J-K_S$ vs. $K_S$ color-magnitude diagram for S208.
The sources detected in both IRAC MIR and NIR bands are shown {with}
squares.
The sources in the cluster region (inside the green circle in
Figure~\ref{fig:disk_dist}) are shown with colored squares, whereas
those outside the cluster region are shown with black open squares.
The different colors of the squares show different disk categories from
the SED slopes: red for class I candidates, magenta for class II
candidates, cyan for evolved disk candidates, and blue for class III
candidates.
The sources identified as cluster members are shown with filled colored
squares, whereas those in the cluster region, which are not identified
as cluster members, are shown with open colored squares.
The sources detected only in the NIR bands are shown with dots.  The
sources in the cluster region are shown with filled dots, whereas those
outside the cluster region are shown with open dots.
The locus of dwarf main sequence stars with spectral types O9 to M6
(corresponding to masses $\sim$20--0.1 $M_\odot$, respectively) from
\citet{Bessell1988} is shown as a black line.
Isochrone models for the age of the S208 cluster, 0.5 Myr old, with
extinctions $A_V = 0$ and 2 mag are shown as thick and thin blue lines,
respectively.
The short arrow shows the reddening vector for $A_V = 2$ mag relative to
the isochrone models with $A_V = 0$ mag for the mass of 20 $M_\odot$.
The isochrone models are from \citet{Lejeune2001} for masses $M/M_\odot
\ge 7$, from \citet{Siess2000} for masses $3 \le M/M_\odot \le 7$, and
from \citet{{D'Antona1997},{D'Antona1998}} for masses $M/M_\odot \le
3$.
The tick marks on the blue lines show the positions of the isochrone
models for the masses 20, 10, 5, 3, 1.5, 1, 0.5, and 0.1 $M_\odot$, with
blue numbers showing the masses.
A 4 kpc distance is assumed for the main-sequence locus and the
isochrone tracks.
The black dashed lines are from isochrone tracks for the mass of 1 and
0.5 $M_\odot$, and parallel to the reddening vector.
The gray dashed line shows the case of 1.5 $M_\odot$ assuming $D=4.8$ kpc. 
The gray solid and dot-dashed lines show the approximated isochrone
models assuming $A_V = 0$ and 4 mag, respectively, which were used to
identify the cluster members in \citet{Yasui2016_S208}.
The dotted lines show the limiting magnitudes in the NIR bands
(10$\sigma$) from \citet{Yasui2016_S208}.}
\label{fig:CM_S208}
\end{center}
\end{figure}

\begin{figure}[!h]
\begin{center}
\includegraphics[scale=0.55]{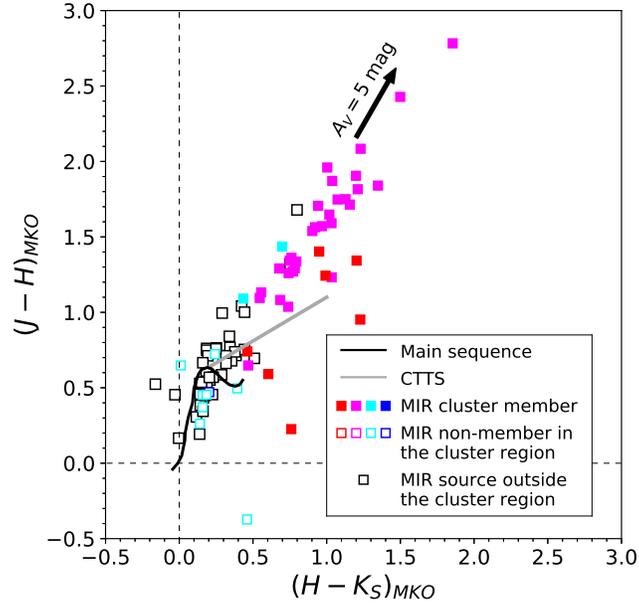} 
\caption{$(H-K_S)$ vs. $(J-H)$ color-color diagram for S208.
The square symbols are the same as those in Figure~\ref{fig:CM_S208}. 
The solid curve in the lower left portion is the locus of points
corresponding to unreddened main-sequence stars.
The classical T Tauri star (CTTS) locus is shown by the gray line.}
\label{fig:CC_S208}
\end{center}
\end{figure}

The point sources detected in the IRAC bands are shown in the $J-K_S$
vs. $K_S$ color-magnitude diagram (Figure~\ref{fig:CM_S208}) using
different symbols to indicate whether a sources is located inside or
outside of the cluster region and to represent the $\alpha_0$ value
classification.
The sources inside the cluster region (inside the green circle in
Figure~\ref{fig:disk_dist}) are shown {with} colored
squares, whereas those outside are shown {with} black
open squares.
The different-colored squares have the same meanings as those in
Figure~\ref{fig:disk_dist}, showing different disk evolutionary stages
as categorized from the SED slopes (i.e., red for class I candidates,
magenta for class II stars, cyan for {evolved disk}
candidates, and blue for class III stars).
ID 44 is shown based on the $\alpha$ value, and ID 59 is not shown in
the figure because it is not detected in the NIR bands.
The figure shows that many of the sources with $A_V \le 2$ were
categorized {into evolved disk} sources, whereas many of
the sources with $A_V \ge 2$ mag were categorized {into}
class I or II objects.
The clear difference suggests that the sources with $A_V \le 2$ mag are
likely to be foreground sources.
Therefore, we identified herein the sources in the cluster region with
$A_V \ge 2$ mag as cluster members.
Figure~\ref{fig:CM_S208} shows the probable cluster members
{with} colored filled squares, whereas sources located in
the cluster region, which are not identified as cluster members, are
shown {with} colored open squares.
Out of 51 IRAC sources located in the cluster region, 10 were excluded
as cluster members because they are probable foreground sources.

To identify cluster members conveniently, \citet{Yasui2016_S208} used
the isochrone models approximated as a straight line (the solid gray
line in Figure~\ref{fig:CM_S208}) because the possible range of distance
and age for S208 were very large at the beginning of their analysis.
They identified as cluster members the sources located to the right of
the gray dot-dashed line in Figure~\ref{fig:CM_S208}, which approximates
isochrone models with extinctions of $A_V = 4$ mag.
All the cluster members identified here (colored filled squares) are
located to the right of the gray dot-dashed line, except for two sources
at $K_S \sim 16$--17 mag that are categorized {into}
class I candidates herein.
Therefore, we confirmed that the method of identifying members here is
basically consistent with that of \citet{Yasui2016_S208}.

Note that one source (ID 59) was not detected in the NIR bands and is
not shown in Figure~\ref{fig:CM_S208}. 
The source was categorized {into} a class I candidate
from its $\alpha$ value.
Class I candidates are not seen outside of the cluster region in
Figure~\ref{fig:disk_dist}; therefore, this source is likely to be a
cluster member.
In summary, we identified 41 sources as probable S208 cluster members.
Table~\ref{tab:S208cl_Spi} labels the cluster members ``CL.''
Sources inside the cluster region, which are not identified as cluster
members, are labeled ``(CL).''


\section{Discussion} \label{sec:discussion_s208}

\subsection{SED Classification} \label{sec:sed_class}

\begin{figure}[!h]
\begin{center}
\vspace{-8em}
\includegraphics[scale=0.8]{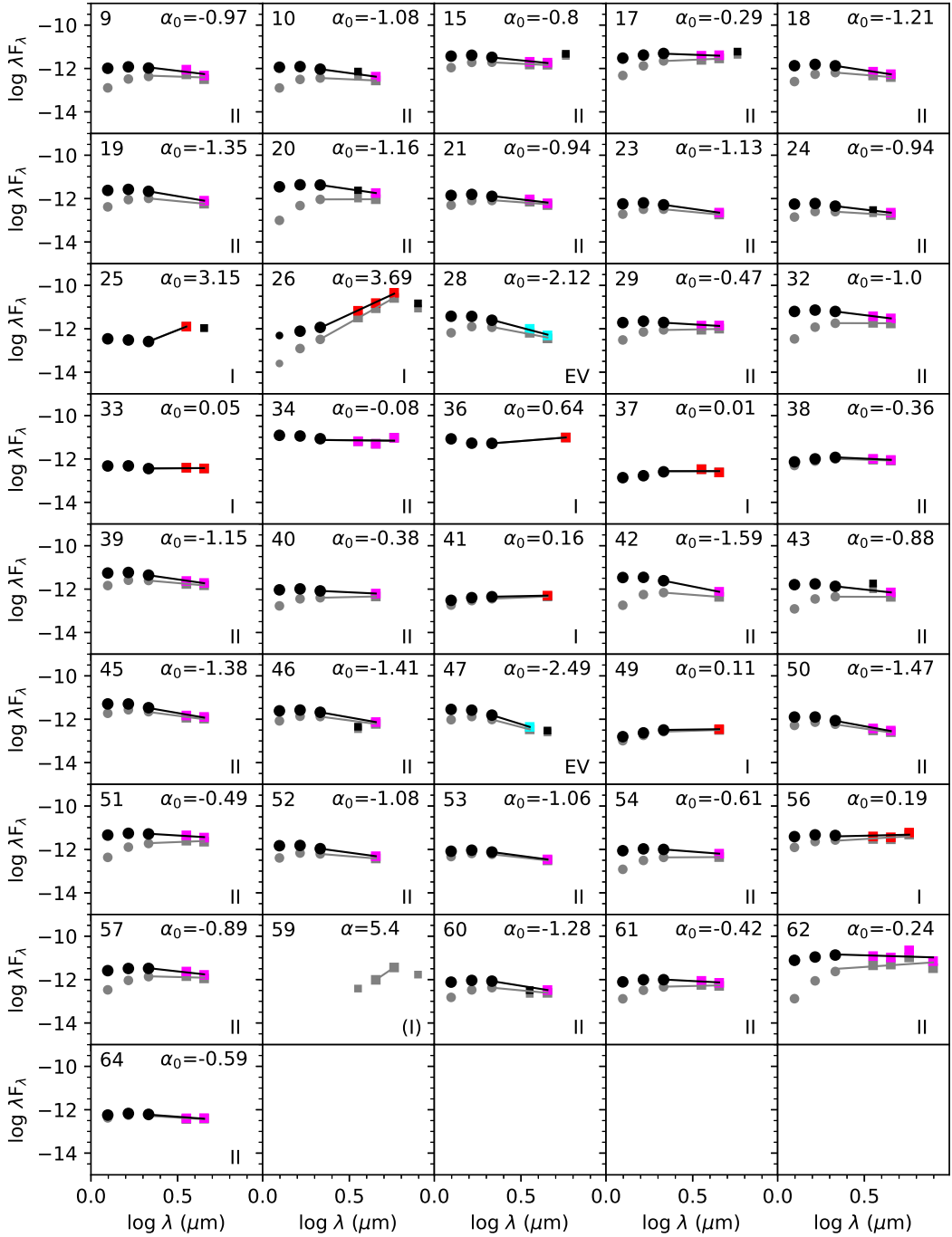}

\vspace{-2em}
 \caption{Spectral energy distributions for the probable cluster members.
Data extending from the NIR {\it J} band (1.26 $\mu$m) to the IRAC
bands are shown for each source.
The IRAC MIR and NIR data are represented by squares and circles,
respectively.
The unit of the y-axis is erg cm$^{-2}$ s$^{-1}$.
The large symbols show detections with $\ge$10$\sigma$, whereas the
small symbols illustrate detections with $\ge$5$\sigma$ but
$<$10$\sigma$.
The observed data are shown in gray, whereas those corrected for
extinction using the $A_V$ derived for each object are shown in black
(NIR data) or in color (MIR data).
The ID and $\alpha_0$ value for each source are shown in the upper left
and upper right corners of each panel, respectively.
The $\alpha$ values are shown for those sources for which $\alpha_0$
values cannot be determined due to {non-detection} in the
NIR bands.
The classification based on the SED slope for each source is shown at
the bottom right of each panel: I, II, EV, and III for class I
candidates, class II stars, stars with evolved disks, and class III
stars, respectively.
The square symbols have different colors based on the SED
classifications of the $\alpha_0$ values: red for class I, magenta for
class II, cyan for stars with evolved disks, and blue for class III
stars.
The results of fits to determine the SED slopes are shown by black and
gray lines for $\alpha_0$ and $\alpha$, respectively.} 
\label{fig:SED_S208}
\end{center}
\end{figure}

Figure~\ref{fig:SED_S208} illustrates the SEDs of the 41 sources
identified as cluster members in Section~\ref{sec:member}.
The data from the NIR {\it J} band (1.26 $\mu$m) to the IRAC bands are
shown for each source.
The IRAC MIR and NIR data are represented as squares and circles, respectively.
Large symbols show detections with $\ge$10$\sigma$, whereas small
symbols depict detections with $\ge$5$\sigma$ but $<$10$\sigma$.
The observed data are shown in gray, and the data corrected for
extinction using the $A_V$ derived for each object are shown in black
for the NIR and in color for the MIR bands.
The ID and the $\alpha_0$ value for each source are shown in the upper
left and upper right of each panel, respectively.
The classification of the disk evolutionary stage based on the
$\alpha_0$ value for each source is shown at the bottom right of each
panel: I, II, EV, and III for class I candidates, class II stars, stars
with evolved disks, and class III stars, respectively.
The square symbols have different colors based on the classification:
red for class I, magenta for class II, cyan for stars with evolved
disks, and blue for class III stars.
For ID 59, the $\alpha$ value and the classification based on the
$\alpha$ value are shown because the $\alpha_0$ values cannot be
estimated.
The results of fitting the data from the NIR {\it K}-band to the MIR
IRAC bands to determine the SED slope are shown with black and gray
lines for $\alpha_0$ and $\alpha$, respectively.

Figure~\ref{fig:cum_DF} shows the cumulative distribution of $\alpha_0$
for the S208 cluster members {with} a black histogram.
Here, we used probable cluster members for which the values of
$\alpha_0$ were estimated, thus excepting ID 59.
In a later section {(Section~\ref{sec:comparison_sed})},
we will compare the distribution with clusters in the solar
neighborhood.
Although the sensitivities must be the same when making comparisons, the
sources detected in at least the 3.6- or 4.5-$\mu$m bands were plotted
because the sensitivities of the 5.8- and 8.0-$\mu$m bands could not be
estimated due to the very small number of detected objects.
For this reason, {ID 36}, which was detected only in the
5.8-$\mu$m {band among the four} IRAC bands, was not
included in this plot.

\subsection{MIR Disk Fraction} \label{sec:disk_fraction}

The disk fraction, which is the frequency of stars with disks within a
young cluster, has been widely studied for various star-forming clusters
in the solar neighborhood.
The disk lifetime is estimated from the disk fractions for various
clusters with assumed ages \citep[e.g.,][]{Haisch2001ApJL}.
Among 41 probable cluster members identified
Section~\ref{sec:results_s208}, nine sources are categorized
{into} class I candidates, 30 sources
{into} class II stars, two sources {into
evolved disk} candidates, and no sources {into} class III
stars.
The classifications are basically based on the $\alpha_0$ values, but
for ID 59, only the $\alpha$ value is available to be used for
classification.

Because the disk fraction is {suggested} to depend on
stellar mass, with larger disk fractions for lower-mass stars
\citep{{Kennedy2009},{Yasui2014},{Ribas2015}}, it is important to
estimate the disk fraction using sources in a limited mass range.
We estimated the disk fraction for sources with masses $\ge$1 $M_\odot$,
which is the {mass detection} limit of this study
(Section~\ref{sec:detect}).
We did not have a spectral type{, and thus mass,} for
each source; therefore, we used the NIR color-magnitude diagram
(Figure~\ref{fig:CM_S208}) to select such sources.
In this diagram, we regard the sources above the black dashed line
corresponding to 1 $M_\odot$ as sources with masses $\ge$1 $M_\odot$.
As shorter wavelengths tend to be unaffected by {disk
excess emission,} it may be better to use the {\it J}-band magnitude
rather than the {\it K}-band magnitude on the y-axis in the
color-magnitude diagram.
However, we used {the} figure that matched on in
{the} previous paper by \citet{Yasui2016_S208} after
confirming that no difference exists between the selected sources when
using these two diagrams.
A total of 27 sources met this condition, and 25 among them were class I
or II objects.
Therefore, the disk fraction for sources with masses $\ge$1 $M_\odot$
was estimated to be 93\% $(25/27)$ using only sources detected in the
IRAC bands.

However, this value may be an upper limit because some sources were not
detected even when the stellar masses inferred from the color-magnitude
diagram were $\ge$1 $M_\odot$.
Because of the lower resolution, stars with no excess are faint and
cannot be detected easily in regions with extended emission and other
bright point sources, so stars with thick disks may possibly be detected
selectively.
In the NIR color-magnitude diagram, sources with masses $\ge$1 $M_\odot$
in the cluster region, which were not detected in the IRAC bands in this
study, are depicted by black dots located above the black dashed line
corresponding to 1 $M_\odot$.
We found 12 such sources.
Assuming that all of these objects were not detected because they did
not have disk excesses, the disk fraction for sources with $\ge$1
$M_\odot$ was 64\% [25/(27 + 12)], which should be a lower limit for the
MIR disk fraction.
Assuming that the magnitudes in the IRAC bands of the sources not
detected in those bands are the corresponding limiting magnitudes
($\simeq$15.5 and $\simeq$15.0 mag in the 3.6- and 4.5-$\mu$m bands,
respectively; see Section~\ref{sec:Spitzer}), only two of the 12 objects
were found to have optically thick disks (two class II stars, two
{evolved disk} stars, and eight class III stars),
suggesting that disk fraction is 69\% [(25+2)/(27+12)].
Therefore, the MIR disk fraction for the S208 cluster is estimated to
lie between 64\% and 93\%, but it is more likely to be closer to 64\%.

We discussed cluster membership in Section~\ref{sec:member}, and we
selected the following as cluster members: sources located in the
cluster region ($r=50\arcsec$) and with extinctions $A_V \ge 2$.
The extinction criteria were used to remove contamination by foreground
sources.
However, some sources may have been removed in this way as foreground
stars even though they are actually cluster members.
ID 44 is thought to be a probable dominant exciting source for the
\ion{H}{2} region, which is suggested to be a late O- or early B-type
star (see Section~2.1 in \citealt{Yasui2016_S208}), and it is possibly a
cluster member.
Ten sources are located in the cluster region but are not identified as
cluster members in this study.
Eight among them are suggested to have stellar masses $\ge$1 $M_\odot$;
one is a class II source and seven are stars with evolved disks.
Furthermore, considering sources with masses $\ge$1 $M_\odot$ that are
detected in the NIR bands but not in the IRAC bands (two sources with
$A_V < 2$ mag along with 12 sources with $A_V \ge 2$ mag), the disk
fraction for all sources in the cluster region is estimated to be 53\%
[(25 + 1)/(27 + 12 + 10)].

In principle, we should also consider background sources in the Galaxy;
however, considering the large {Galactrocentric} distance
of S208 ($R_G = 12$ kpc), the number of the {Galactic}
background stars is expected to be very small.
In addition, CO emission was detected in S208 by \citet{Blitz1982}.
The H2 column density was also estimated to be very large: $\sim$4--$5
\times 10^{22}$ cm$^{-2}$, which corresponds to $A_V = 50$ mag
\citep{Yasui2016_S208}.
It may be difficult to see beyond the CO molecular cloud due to this
very high column density.
Moreover, the effect of the background sources should be negligible.
\citet{Gutermuth2009} pointed out that contamination can also arise from
non-YSO extragalactic sources with excess infrared emission, including
star-forming galaxies, broad-line active galactic nuclei (AGNs), and
unresolved knots of shock emission from outflows that collide with the
cold cloud material.
As an example, \citet{Gutermuth2008} identified 44 sources that are
likely to be star-forming galaxies and AGNs in a field of
$\sim$$40\arcmin \times 30\arcmin$.
Because the region of the S208 cluster is estimated to be a region with
a 50$\arcsec$ radius, the number of such sources is estimated to be
almost zero, and its effect on the disk fraction should thus be
negligible.

In summary, the disk fraction for sources with masses $\ge$1 $M_\odot$
is estimated to be between 64\% and 93\%, taking into account sources
that were not detected in the IRAC bands.
Even considering various uncertainties (the possible existence of actual
cluster members removed as foreground sources and that of
{Galactic} and extragalactic background sources), the
disk fraction is estimated to be at least 53\%.

\subsection{Comparison with Clusters in the Solar Neighborhood}
\label{sec:comparison_solar}

This section compares the properties of the MIR disks in S208
(Sections~\ref{sec:sed_class} and \ref{sec:disk_fraction}) with those in
nearby clusters.
The sources detected in the S208 MIR observations discussed herein are
basically intermediate-mass stars.
We compared the results obtained here to those of intermediate-mass
stars in the solar neighborhood.
Although the {mass detection} limit was $\sim$1
$M_\odot$, because studies dealing with low-mass stars have generally
been observed down to a stellar mass of $\sim$0.1 $M_\odot$
\citep{Gutermuth2009}, intermediate-mass stars should be appropriate for
a fair comparison.

\subsubsection{Comparison of SED Slopes} 
\label{sec:comparison_sed}

First, we compared the distributions of the intrinsic SED slopes
($\alpha_0$) for the S208 cluster and those of the clusters in the solar
neighborhood.
We selected clusters in the solar neighborhood ($D \le 1$ kpc) that have
the same age as S208 {($\le$1 Myr old)} because the
distributions of $\alpha_0$ are known to change with age
\citep{Hernandez2007}.
We selected two clusters, NGC 2024 (0.3 Myr) and NGC 1333 (1 Myr), which
were discussed in previous studies of disk fraction/disk evolution
(e.g., \citealt{Haisch2001ApJL}, \citealt{Hernandez2008}) and about
which papers have been published on the data taken by the {\it
Spitzer}/IRAC (\citealt{Megeath2012} for NGC 2024 and
\citealt{Gutermuth2008} for NGC 1333).
In those references, only the values of the observed SED slopes
($\alpha$) were shown instead of the $\alpha_0$ values.
The $\alpha$ values tend to be estimated as higher than the $\alpha_0$
values, as pointed out by \citet{Gutermuth2008}; therefore, we estimated
the $\alpha_0$ value for each source using the MIR IRAC and NIR {\it
JHK}-band data for these two clusters following the same procedure as in
this paper.
Because of the difference in distance modulus between the nearby
clusters ($D=415$ pc for NGC 2024 and 250 pc for NGC 1333) and S208
($D=4$ kpc), 4.9 and 6.0 mag, respectively, we used only the data up to
10.6 mag $(=15.5-4.9)$ and 9.5 mag $(=15.5-6.0)$ in the 3.6-$\mu$m band
and up to 10.1 mag $(=15.0-4.9)$ and 9.0 mag $(=15.0-6.0)$ in the
4.5-$\mu$m band for NGC 2024 and NGC 1333, respectively, to match
sensitivities.
We used data only up to the 4.5-$\mu$m band because the numbers of
cluster members in S208 detected at 5.8 and 8.0 $\mu$m were very small.
For the NGC 2024 data, we used sources in the same field as
\citet{Haisch2000}, who presented the results of {\it L}-band imaging of
the cluster, centered at
$\alpha_{2000} = 5^{\rm h} 41^{\rm m} 45.27^{\rm s}$, 
$\delta_{2000} = -1^\circ 54\arcmin 31.49\arcsec$, 
with a $10.5\arcmin \times 10.5\arcmin$ field of view.

The cumulative distributions of $\alpha_0$ obtained for NGC 2024 and NGC
1333 are shown as red and blue histograms, respectively, in
Figure~\ref{fig:cum_DF}.
Although the distributions of both nearby clusters are very similar, the
distributions appear to take {smaller values} overall
compared to the distribution for the S208 cluster.
This may be caused by the different environments, i.e., metallicity.
However, as discussed in Section~\ref{sec:disk_fraction}, if we take
into account {the} sources that were identified as
members based on NIR observations but were not detected in the IRAC
bands, and if we assume that their magnitudes in the IRAC bands are the
corresponding limiting magnitudes, the distribution of the S208 cluster
is the histogram shown by the black dashed line.
Although the distribution still appears to take slightly larger values
than the distributions of the nearby clusters, no significant
differences are present, judging from a K-S test.

\subsubsection{Comparison of the MIR Disk Fractions} 

We then compared the MIR disk fractions.
Disk fractions for clusters are often used to derive the disk lifetime
by plotting them as a function of the cluster age.
{Disk fraction--age plots} were first proposed for
low-mass stars using NIR {\it JHK}-band and {\it L}-band data
\citep{{Lada1999}, {Haisch2001ApJL}}.
From the results that the disk fractions for many of the star-forming
clusters were $\sim$100\% at age $\sim$0, and that the values decreased
on a timescale of $\sim$5--10 Myr as the age increased, the disk
lifetime was estimated to be approximately 5--10 Myr.
Similar plots were then generated from the {\it Spitzer} MIR data, and
the obtained disk fraction{, and thus the disk lifetime,}
was found to be comparable to the previous derivation
\citep[e.g.,][]{Hernandez2008}.
Subsequently, it was pointed out that a mass dependence exists in the
disk lifetime derived using NIR and MIR wavelengths
\citep{{Hernandez2005}, {Kennedy2009}, {Yasui2014}, {Ribas2015}},
higher-mass stars have shorter lifetimes.
Moreover, recent results from ALMA submillimeter continuum observations
\citep{Ansdell2017} show the opposite trend, with submillimeter
continuum emission weakening with age for lower-mass stars, indicating
that the mass of the dust disk decreases more rapidly for lower-mass
stars.
However, it should be noted that the sample of intermediate-mass stars
is very small, the sample of intermediate-mass stars is not always
detected, and there is a very large variation in the derived masses
within each cluster.

In Figure~\ref{fig:DF_age}, we plotted the MIR disk fractions for the
intermediate-mass stars (MIR IMDFs) for the nearby clusters, shown with
large circles.
The plots are from \citet{Yasui2014}, who selected their target clusters
from previous studies of the disk fraction/disk evolution
\citep[e.g.,][]{{Haisch2001ApJL},{Hernandez2005},{Hernandez2008},
{Kennedy2009}} and derived the MIR IMDF using stars with masses 1.5--7
$M_\odot$ selected based on spectral types from the literature for the
same clusters as in previous studies.
The MIR IMDF evolution curves are also represented by a thick curve from
\citet{Yasui2014}.
In this plot, the IMDFs for many of the star-forming clusters were
$\sim$100\% at age $\sim$0, and the values decreased as the ages
increased.
We also plotted the disk fraction obtained for S208, shown
{with} a red square with the estimated uncertainties.
For S208, which is {suggested} to have a very young age
($\sim$0.5 Myr; \citealt{Yasui2016_S208}; see Section~\ref{sec:intro}),
the disk fraction estimated herein has a relatively high value, between
64\% and 93\%.
This is consistent with the results from clusters in the solar
neighborhood.
Although the disk fraction for the S208 cluster has a lower limit of
53\% even considering the uncertainties, we still did not observe a
significant difference between the disk fraction obtained for S208 and
the values for clusters in the solar neighborhood.

\begin{figure}[!h]
\begin{center}
 \includegraphics[scale=0.5]{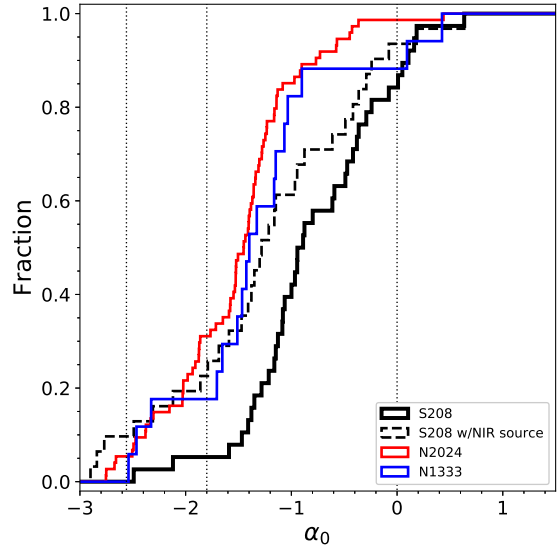}
\caption{Cumulative distributions of the intrinsic SED slopes
($\alpha_0$) for S208, NGC 2024, and NGC 1333 shown by the black, red,
 and blue histograms, respectively.
For S208, the distribution including only the stars detected in the IRAC
bands is shown by the thick black line, and the distribution including
the stars detected in NIR bands is shown by the black dashed line.
 The vertical dotted lines show the borders of the $\alpha_0$ values
 among class I stars, class II stars, stars with evolved disks, and
 class III stars.} 
\label{fig:cum_DF} 
\end{center}
\end{figure}

\begin{figure}[!h]
\begin{center}
\includegraphics[scale=0.5]{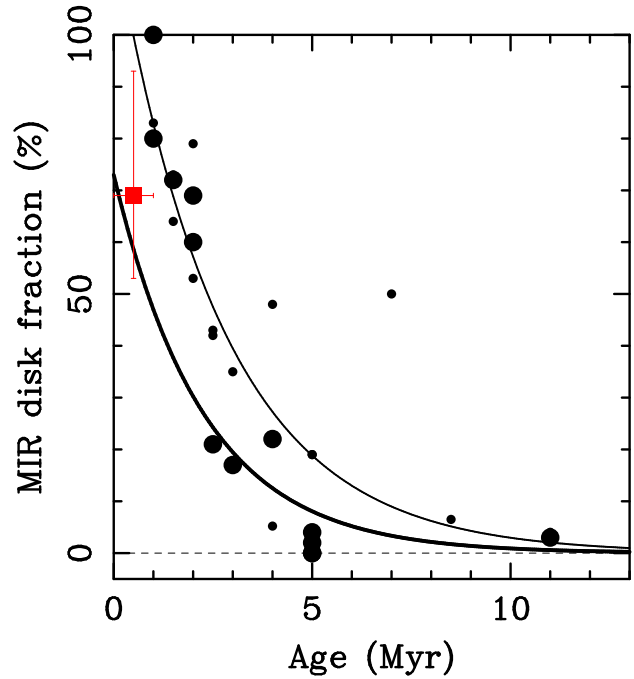}
 \caption{MIR disk fraction as a function of cluster age.
The MIR disk fraction for the S208 cluster is shown by
{a} red filled square, whereas those of young clusters
with solar metallicity are shown by black filled circles.
The large circles are for intermediate-mass stars, whereas the small
circles are for low-mass stars.
The thick and thin black curves show the evolution of the disk fraction
at solar metallicity for intermediate- and low-mass stars, respectively.
All the data plots for clusters and the {disk fraction} evolutionary
curves at solar metallicity are from \citet{Yasui2014}, who derived the
disk fraction for intermediate-mass stars using data for stars with
1.5--7 $M_\odot$ from previous studies.
They also collected the results for disk fractions for low-mass stars
from previous studies (see the text).}
\label{fig:DF_age}
\end{center}
\end{figure}

A few points must be noted.
First, various criteria have been used as the boundary values employed to determine whether or not a star has a disk:
$\alpha_0 =-1.8$ by \citet{Hernandez2007};
$\alpha_0 = -2.0$ by \citealt{Lada2006};
and $\alpha _0= -2.2$ by \citet{Kennedy2009} and \citet{Yasui2014}.
We used $\alpha_0 = -1.8$ for S208 in this study.
Although the disk fraction became even larger when the boundary values
$\alpha_0 = -2.0$ or $-$2.2 were used, this change does not
significantly affect the discussion here.
Second, the SED slopes derived using shorter wavelengths tend to be
lower than those derived using longer wavelengths.
The plots for clusters with solar metallicity used herein are from
\citet{Yasui2014}, which used data from references that mainly extended
up to 8.0-$\mu$m wavelength.
Moreover, the SED slopes have generally derived from the MIR data alone
(i.e., without the NIR data) in MIR studies of nearby clusters
\citep[e.g.,][]{Lada2006}.
In contrast, the SED slopes for the S208 cluster members are derived
from NIR {\it K}-band data and MIR data up to 4.5 $\mu$m.
We derived $\alpha_0$ values only from the IRAC bands (3.6--8.0 $\mu$m)
for detections at more than at 10$\sigma$ {(hereafter,
$\alpha_{0 {\rm (MIR)}}$)} for the sources in NGC 2024 and NGC 1333 for
which the values of $\alpha_0$ were derived from the NIR {\it K}-band to
the IRAC 4.5-$\mu$m band and were used in plotting the cumulative
distributions shown in Figure~\ref{fig:cum_DF}.
We then compared the $\alpha_0$ values for the sources for which values
were obtained by either derivation.
Consequently, $\alpha_{0 {\rm (MIR)}}$ is larger by $-$0.04 $\pm$ 0.4
mag for NGC 2024 and by $+$0.04 $\pm$ 0.5 mag for NGC 1333, suggesting
that the systematic differences are very small although the dispersions
are relatively large.
Although the SED slopes for the S208 cluster members may be slightly
higher{, if they can be observed with sufficient
sensitivity up to 8.0-$\mu$m wavelength,} the change in the disk
fraction should be quite small, judging from the cumulative distribution
of $\alpha_0$ for S208 shown in Figure~\ref{fig:cum_DF}.
Third, some low-mass stars were included among the S208 cluster members,
although the majority were intermediate-mass stars.
The disk fraction is {suggested} to depend on the stellar
mass, with larger disk fractions for the lower-mass stars
\citep{{Kennedy2009},{Yasui2014},{Ribas2015}}.
%
Thus, for comparison, we also plotted the MIR disk fraction for the
low-mass stars (MIR LMDFs) shown by the small circles in
Figure~\ref{fig:DF_age} and the evolution curve for the MIR LMDFs shown
by a thin curve from the same reference as for IMDF.
The point representing S208 falls between the thick and thin lines,
suggesting that the disk fraction for the S208 cluster is comparable to
those for the nearby clusters. In any case, these three points did not
have a big impact on the results herein.



\subsection{Comparison with Results in Other Low-metallicity Environments} 
\label{sec:LowMeta}

Some previous studies focused on the evolution of protoplanetary disks
in star-forming regions in low-metallicity environments.
In a series of papers, we compiled a list of 
low-metallicity (${\rm [O/H]} \simeq -1$ dex) young clusters in the
Galaxy and presented deep NIR imaging observations with a detection
limit of $\sim$0.1 $M_\odot$.
The estimated disk fractions were relatively high for very young
clusters, but they declined rapidly on the timescale
of $\sim$1 Myr,
suggesting that the lifetime of protoplanetary disks in such
environments is shorter than that in clusters with solar metallicity
\citep{{Yasui2016_S208}, {Yasui2009}, {Yasui2010}, {Yasui2016_S207},
{Yasui2021}}.
Although the results for the MIR disk fraction of the S208 cluster
discussed herein appears to be contrary to our
previous results,
the difference can be attributed to the difference of the stellar mass
range.  Our previous results are for stars down to low-mass stars of
$\sim$0.1 $M_\odot$, thus primarily for low-mass
stars, while
the results herein are for stars down to $\sim$1 $M_\odot$, thus
primarily for intermediate-mass stars.
Disk properties could be quite 
different because stars with different masses have different
evolutionary processes with different timescales.
We will discuss this point in the next subsection.

Another example in the Galaxy is S284
located at $D\sim 4$ kpc and
in a possibly low-metallicity environment.
%
\citet{Cusano2011} {found} that a significant fraction
 of {the} YSOs {in Dolidze 25 cluster in
 S284}
still {have} disks {judging from} the
 optical spectroscopic data{,} including the H$\alpha$
 band{,} and {from} the photometric
 data{, including} the {\it Spitzer}/IRAC band.
\citet{Kalari2015} {also found} that no evidence
{to support} a systematic change in the mass accretion
rate with metallicity {from optical spectroscopic study
of Dolidze 25.}
Both results deal with the identical mass range of stars in this paper
($\sim$1--2 $M_\odot$){,} and {the
consistency with the results for the S208 cluster discussed herein
appears to be a natural consequence}.
However, {care should be taken as to whether this
region is an object that can be directly compared to S208.
For example, the metallicity of this region is still uncertain: 
{the metallicity of S284 is reported from a low of
$-$0.9--$-$0.6 dex \citep{Lennon1990}
to a high of $-$0.3 dex \citep{Rudolph2006},}
which is more like close to solar metallicity. 
{Furthermore, a complex star formation history 
has been pointed out in this region:}}
{there are} two populations {of different ages}
\citep{Delgado2010}{, and} sequential
star-forming activities occur in {this} \ion{H}{2}
region {over} a large region
{that extends over} $\sim$20 pc
{\citep{Puga2009}}.
Therefore, {it may be difficult to consider} disk
properties {at} the level of individual clusters
(spatial scale{s} of $\sim$1 pc), such as the S208
cluster (Section~\ref{sec:member}) and the clusters in the solar
neighborhood \citep{Adams2006}.

Other possible comparison targets are the 
star-forming regions in the LMC/SMC, 
which have similarly low metallicity (${\rm [O/H]} \simeq -0.3$ dex and 
$\simeq -0.7$ dex for the LMC  and SMC, respectively; 
e.g., \citealt{Dufour1982}) as those in the outer Galaxy. 
%
%
Studies aimed at dust disks
ranged from {\it L}-band imaging from the South Pole
\citep{Maercker2005} 
to {\it Spitzer} surveys called SAGE-LMC and SAGE-SMC 
using {\it Spitzer} IRAC  (3.6, 4.5, 5.8, and 8.0
 $\mu$m) and MIPS (24, 70, and 160 $\mu$m) 
 \citep{{Meixner2006},{Gordon2011}}, and
 they identified a very large number of new YSO candidates. 
However,  due to the lower spatial resolutions and higher mass detection limits 
compared to the Galactic studies 
($\simeq$0.5 pc in spatial resolution and  stellar masses $\gtrsim$5 $M_\odot$ 
even using {\it Spitzer} \citep[e.g.,][]{Romita2010}), 
these studies appear to result in selective detection of stars with very early stage disks
with higher masses \citep{Whitney2008}, 
so making a direct comparison in difficult. 
As for gas accretion disks, 
\citet{De Marchi2010} and their group's work
suggested that the disk lifetime in LMC/SMC
with stellar masses $\gtrsim$0.5 $M_\odot$
is approximately twice as long as that in solar-metallicity environments
from {\it HST} optical imaging including the H$\alpha$ band. 
The results  seem to be inconsistent with the results here 
for the MIR dust disks for intermediate-mass stars, which might suggest 
that factors other than metallicity have also influenced the
protoplanetary disk evolution.
However, considering that the complexity of the star formation history
in the LMC and SMC has also been pointed out (e.g.,
\citealt{Harris2009}) and that those observations dealt with very large
spatial scales (e.g., $\sim$$40 \times 40$ pc in \citealt{De
Marchi2010}), a direct comparison with our results in a single
star-forming area is somewhat difficult.  

In summary, {comparisons} with Galactic
studies {show that} our results are consistent with those
for the same mass range, {i.e.,} intermediate-mass
stars, but {are} inconsistent with those for
{a} different mass range,
{i.e.,} low-mass stars. This may be because stars with
different masses have different evolutionary processes with different
timescales, which can make disk properties different.
{As for comparisons} with extragalactic studies,
{the LMC and SMC have been studied mainly,
and it seems difficult to make direct comparisons with this study
because the spatial resolution, mass detection limit, and observed
spatial scales are very different.}
Although sequential star formation has been suggested in a large scale
around S208, the S208 cluster is a single cluster located in a single
\ion{H}{2} region with {a} cluster size
{comparable} to those in the solar neighborhood ($\sim$1
pc in radius){. This suggests} that the star-forming
history {may} relatively simple compared to
{that in} other regions studied in low-metallicity
environments, {for which the} star formation histories
are suggested to be complex.
Therefore, S208 should be a very good laboratory for exploring
protoplanetary disk evolution by observing individual star-forming
cluster{s at the same} scale in different metallicity
environments, {thus enabling} the investigation of the
impact of protoplanetary disk evolution due {purely} to
metallicity change{s} down to $\sim$1 $M_\odot$-mass
stars{,} based on MIR data for the first time.


\subsection{Implications for the Evolution of Protoplanetary Disks Surrounding Intermediate-mass Stars} \label{sec:diskevolve}

We compared the cumulative $\alpha_0$ distributions for the S208 cluster
to those for nearby clusters with approximately the same age in
Section~\ref{sec:comparison_solar}, and we found that the distribution
for the S208 cluster appears to take a larger value overall compared to
the distributions for nearby clusters with approximately the same age,
where data of the same sensitivity were used.
The SED slopes are known to indicate the degree of dust growth/settling
\citep{{D'Alessio2006},{Hernandez2007}}; thus, this result may suggest
that the degree of dust growth/settling in the MIR disks of the S208
cluster members is low compared to those of the other cluster members in
the solar neighborhood.
This is consistent with the idea that the probability of dust collision
is low due to the low dust density in such environments.
However, if we also consider {the} sources that are
identified as cluster members based on NIR observations, which are not
detected in the MIR bands, no significant difference is found,
suggesting that the degree of dust growth/settling in protoplanetary
disks does not significantly change with metallicity differences of
$\sim$1 dex, at least until this age ($\sim$0.5 Myr).

In this study, we also found no difference in the MIR disk fractions for
intermediate-mass stars between the S208 cluster and clusters in the
solar neighborhood.  This may suggest that {disk
dispersal} mechanisms that work effectively in intermediate-mass stars
have either no, or else a very weak, dependence on metallicity.
For example, metallicity-independent photoevaporation, such as
extreme-ultraviolet (EUV) photoevaporation, may be dominant.
It may also be possible that metallicity-dependent photoevaporation
(e.g., FUV and X-ray photoevaporation) is less dependent on metallicity
than currently thought \citep{{Nakatani2018},{Ercolano2010}}.
In any case, the results here indicate only that the disks have not yet
disappeared and that effective mass accretion and/or
photoevaporation, as suggested from the NIR disks for the low-mass stars
in low-metallicity environments, does not seem to occur.

In the first place, it is pointed out that photoevaporation does not
work effectively for stars that have high-mass accretion rates and
accompanying stellar/disk winds because the radiation can be absorbed
before the disk surfaces are reached \citep{Hollenbach2009}, although
the radiation driving photoevaporation for intermediate-mass stars is
expected to be higher than that from low-mass stars \citep{Gorti2009}.
This suggests that the accretion rates need to decrease to
$\sim$10$^{-7}$ $M_\odot$ yr$^{-1}$ for FUV/{hard} X-ray
photoevaporation and to {$\sim$10$^{-8}$ $M_\odot$
yr$^{-1}$} for EUV{/soft X-ray} photoevaporation
\citep{Hollenbach2009}.
The {mass accretion} rates for stars with approximately
the same age as the S208 cluster (0.5 Myr) are estimated to be very high
at solar metallicity (i.e., $\gtrsim$10$^{-7}$ $M_\odot$ yr$^{-1}$ for
low-mass stars).
The {mass accretion} rate {($\dot{M}$)} is
known to have a {stellar mass} dependence, $\dot{M}
\propto M_*^2$ \citep{Hartmann2006}; thus, the {mass
accretion} rates for intermediate-mass stars are expected to be
{even higher.}
Therefore, if all the results obtained at solar metallicity hold for
S208, this suggests that the photoevaporation processes have not yet
started to work in S208, at least for the intermediate-mass stars.

As for the mass accretion suggested by \citet{Yasui2009} as another
possible factor for faster disk dispersal for low-mass stars in
low-metallicity environments, the present results suggest that no strong
dependence on metallicity is observed.
This is consistent with the results for S284 \citep{Kalari2015}, which
is located in a possible low-metallicity environment in the Galaxy
(Section~\ref{sec:LowMeta}).
However, future observations of older star-forming clusters in
low-metallicity environments will ultimately reveal whether the disk
{lifetime} of intermediate-mass stars
{does} not really have a metallicity dependence.
For example, if the disk fractions for such clusters are estimated to be
lower than those in the solar neighborhood, the results here only
indicate that a large fraction of stars have disks in the first stage,
although there is a dependence of metallicity also for intermediate-mass
stars.
A similar result was seen in a very young star-forming region
($\sim$0.1--0.5 Myr) in a low-metallicity environment (i.e., Sh 2-127)
for the NIR disk fraction of low-mass stars, which show a high disk
fraction \citep{Yasui2021}.

This study showed that most sensitive MIR observations with existing
instruments can detect sources located beyond the solar neighborhood
down to approximately the solar mass.
However, {more sensitive} observations that can detect
sources with stellar masses of $\sim$0.1 $M_\odot$ will be necessary in
the future for a comprehensive comparison of
{protoplanetary disk} evolution down to low-mass stars
between low-metallicity and solar-metallicity environments.
Radiation from the central star, which is directly connected to the
photoevaporation rate, varies greatly, depending on the stellar mass and
its stage of evolution; thus, low-mass stars may have completely
different results.
Moreover, observations up to longer wavelengths are important because
longer wavelengths trace the outer part of the protoplanetary disks, and
the results may differ depending on the wavelength used for the
observations (\citealt{Ribas2014}; in the case of nearby clusters).
The forthcoming JWST (James Webb Space Telescope) is planned to enable
such {more sensitive} observations at longer wavelengths.
At the distance of S208 ($D \sim 4$ kpc), stars with stellar masses
$\sim$0.1 $M_\odot$ can be detected with 5$\sigma$ detection capability,
even in the {12-$\mu$m} band that corresponds to the
longest wavelength of {\it Spitzer}/IRAC \citep{Yasui2020}.
For reference, the stellar {mass detection} limit in the
case of the LMC/SMC can only reach $\sim$2--3 $M_\odot$, even with the
sensitivities of JWST.



\section{Conclusions}

The results of a sensitive MIR imaging study of the S208 cluster with
{\it Spitzer}/IRAC were presented herein.
The S208 cluster is a very young ($\sim$0.5 Myr) star-forming cluster
located in an \ion{H}{2} region, Sh 2-208, which is one of the
lowest-metallicity \ion{H}{2} regions in the Galaxy, ${\rm [O/H]} =
-0.8$ dex.
We performed MIR studies at the {individual cluster}
scale (cluster size of $\sim$1 pc) in this low-metallicity environment
to investigate the metallicity dependence of disk evolution by comparing
the results with those obtained for clusters in the solar neighborhood.
The following conclusions were drawn from the MIR photometry in
combination with NIR photometry obtained from NIR imaging with high
sensitivity and spatial resolution from the literature
\citep{Yasui2016_S208}:

\begin{enumerate}
 \item It has been suggested that S208 is located in a
       {sequential}
      star-forming region. From recently released astrometric data from
      Gaia EDR3 for the open clusters in this star-forming region,
      including S208, in combination with information about the distance
      from the literature, a distance $D\sim 4$ kpc {for}
       S208 is suggested
      to be most likely.

 \item We conducted photometry of the IRAC MIR images obtained from the
       {\it Spitzer} archival data in a field of $\sim$$3.5 \times 4$
       arcmin, which is the same area as used for our previous NIR
       study.
       Consequently, 96 sources were detected in at least one MIR band
       at more than 10$\sigma$, covering intermediate-mass stars
       ($\sim$1.5--10 $M_\odot$).
       The {mass detection} limit was estimated to be $\sim$1 $M_\odot$
       for the sources detected in at least one of the IRAC bands;
       however, most sources were detected only in the 3.6- and/or
       4.5-$\mu$m bands.
       The SED slopes were derived from the {\it K}-band to the IRAC
       bands for all the sources detected in at least one IRAC band.
       From the spatial distribution of the SED slopes, a circle of
       radius 50$\arcsec$ was defined as the cluster region, and 51
       sources are located in this region.
       Among these 51 sources, 10 with $A_V \lesssim 2.0$ mag are
       regarded as foreground sources, and 41 with extinctions $A_V
       \gtrsim 2.0$ mag are defined as probable cluster members.

 \item The cumulative $\alpha_0$ distributions for the S208 cluster were
       compared with those in the solar neighborhood having
       approximately the same age.
       The distribution for the S208 cluster appears to take a larger
       value overall compared to the distributions for the nearby
       clusters, which may suggest that dust settling/growth may not be
       occurring efficiently in low-metallicity environments.
       However, if we also consider the sources that are identified as
       cluster members based on NIR observations but are not detected in
       the MIR bands, no significant difference was found, suggesting
       that the degree of dust growth/settling in protoplanetary disks
       does not change significantly with metallicity differences
       $\sim$1 dex, at least until this age ($\sim$0.5 Myr).

\item The MIR disk fraction for the S208 cluster with stellar masses
       $\ge$1 $M_\odot$ was estimated to be between 64\% and 93\%,
       taking into account {non-detected} sources in the
       IRAC bands. Even considering the possible existence of foreground
       and background sources, the MIR disk fraction is estimated to be
       at least 53\%. This value is comparable to the MIR disk fractions
       for clusters in the solar neighborhood with approximately the
       same age.

 \item The results in this study were compared to those previously
       obtained by other studies of low-metallicity regions. By
       comparison with Galactic studies, our results are consistent with
       those for the same mass range{, intermediate-mass
       stars,} but are inconsistent with those for a different mass
       range, low-mass stars. This may be because stars with different
       masses have different evolutionary processes with different
       timescales, which can make the disk properties different. It
       generally seems difficult to make direct comparisons with
       extragalactic studies, for which the LMC and SMC have been
       studied mainly, due to the much larger spatial separations
       ($\sim$0.5 pc), {higher mass detection limits},
       and much larger observed spatial scales for the LMC/SMC.

 \item The result that the MIR disk fraction for S208 is comparable to
       those in clusters with solar metallicity may suggest that
      {disk dispersal} mechanisms that work effectively
       for intermediate-mass stars either have no, or at best a very
       weak, dependence on metallicity. However, this may only indicate
       that a large fraction of stars have disks in the first stage and
       that the {disk dispersal} process (e.g.,
       photoevaporation) does not yet work effectively at this
       stage. Future MIR observations of older star-forming clusters in
       low-metallicity environments are necessary to determine whether
       or not disk evolution has a metallicity dependence for
       intermediate-mass stars.

 \item Using the most sensitive MIR imager, {\it Spitzer}/IRAC, stars
       down to approximately the solar mass can be detected in regions
       located beyond the solar neighborhood. Furthermore, the MIR disk
       properties in such a cluster can be studied at the scale of an
       individual cluster. However, more direct comparisons with studies
       in the solar neighborhood extending down to low-mass stars are
       necessary to derive the comprehensive metallicity dependence of
       disk evolution. For this purpose, stars with stellar masses down
       to $\sim$0.1 $M_\odot$ and MIR wavelengths as long as $\sim$10
       $\mu$m must be studied using next-generation MIR telescopes, such
       as JWST.

\end{enumerate}

\acknowledgments

We are grateful to Prof. Masao Saito for helpful discussions and comments
on the manuscript.
C.Y. is supported by KAKENHI (18H05441) Grant-in-Aid for Scientific
Research on Innovative Areas.
This work is based on observations made with the {\it Spitzer Space
Telescope}, obtained from the NASA/IPAC Infrared Science Archive, both
of which are operated by the Jet Propulsion Laboratory, California
Institute of Technology under a contract with the National Aeronautics
and Space Administration.



{\it Facilities:} {\it Spitzer Space Telescope}, Subaru Telescope,
Gaia.

{\it Software:} IRAF \citep{Tody1993}, MATPLOTLIB, a PYTHON library for
publication quality graphics \citep{Hunter2007}, and NUMPY \citep{van
der Walt2011}.



\clearpage

\begin{longrotatetable}
\startlongtable
\begin{deluxetable*}{lllrrrrrrrrrlll}
 \tablecaption{{\it Spitzer}/IRAC Point-source Catalog
of S208.}
\label{tab:S208cl_Spi}
\tablewidth{700pt}
\tabletypesize{\scriptsize}
\tablehead{
\colhead{ID} & \colhead{R.A. {(J2000.0)}} & 
\colhead{Decl. {(J2000.0)}} &
\colhead{\it J} & 
\colhead{\it H} & 
\colhead{$K_S$} & 
\colhead{3.6 $\mu$m} & 
\colhead{4.5 $\mu$m} &
\colhead{5.8 $\mu$m} & 
\colhead{8.0 $\mu$m} &
\colhead{$A_V$} &
\colhead{$\alpha$} &
\colhead{$\alpha_0$}
& \colhead{Notes}  \\
\colhead{} &
\colhead{(deg)} & 
\colhead{(deg)} & 
\colhead{(mag)} & 
\colhead{(mag)} & 
\colhead{(mag)} & 
\colhead{(mag)} & 
\colhead{(mag)} & 
\colhead{(mag)} & 
\colhead{(mag)} & 
\colhead{(mag)} & 
\colhead{} &
\colhead{} &
\colhead{}  \\
 \colhead{(1)} & \colhead{(2)} & \colhead{(3)} & \colhead{(4)} & 
\colhead{(5)} & \colhead{(6)} & \colhead{(7)} & \colhead{(8)} &
\colhead{(9)} & \colhead{(10)} &
\colhead{(11)} & \colhead{(12)}& \colhead{(13)}
& \colhead{(14)}
} 

\startdata
  1 & 64.838656  & 52.986661  &  16.17 ( 0.01 )  &  15.60 ( 0.01 )  &  15.40 ( 0.01 )  &  15.53 ( 0.09 )  &  ... (...)  &  ... (...)  &  ... (...)  & 0.0  & -2.97  & -2.97  &  \\
2 & 64.857829  & 52.978731  &  14.89 ( 0.01 )  &  14.16 ( 0.01 )  &  13.95 ( 0.01 )  &  14.01 ( 0.07 )  &  13.91 ( 0.08 )  &  ... (...)  &  ... (...)  & 1.1  & -2.76  & -2.86  &  \\
3 & 64.861019  & 53.000657  &  16.89 ( 0.01 )  &  16.17 ( 0.01 )  &  15.79 ( 0.01 )  &  15.50 ( 0.07 )  &  ... (...)  &  ... (...)  &  ... (...)  & 0.0  & -2.21  & -2.21  &  \\
4 & 64.867286  & 53.000356  &  16.24 ( 0.01 )  &  15.56 ( 0.01 )  &  15.20 ( 0.01 )  &  15.05 ( 0.08 )  &  ... (...)  &  ... (...)  &  ... (...)  & 0.0  & -2.44  & -2.44  &  \\
5 & 64.870121  & 53.001991  &  13.55 ( 0.01 )  &  12.56 ( 0.01 )  &  12.27 ( 0.01 )  &  11.79 ( 0.03 )  &  11.73 ( 0.03 )  &  10.89 ( 0.08 )  &  ... (...)  & 4.0  & -1.64  & -1.92  &  \\
6 & 64.870320  & 52.973139  &  15.42 ( 0.01 )  &  14.95 ( 0.01 )  &  14.75 ( 0.01 )  &  14.86 ( 0.08 )  &  14.63 ( 0.08 )  &  ... (...)  &  ... (...)  & 0.0  & -2.68  & -2.68  &  (CL)\\
7 & 64.870955  & 52.988014  &  13.95 ( 0.01 )  &  13.30 ( 0.01 )  &  13.29 ( 0.01 )  &  12.92 ( 0.05 )  &  12.85 ( 0.06 )  &  ... (...)  &  ... (...)  & 1.6  & -2.21  & -2.35  &  (CL)\\
8 & 64.871173  & 52.964897  &  15.16 ( 0.01 )  &  13.48 ( 0.01 )  &  12.69 ( 0.01 )  &  10.94 ( 0.03 )  &  10.14 ( 0.03 )  &  8.59 ( 0.03 )  &  8.59 ( 0.03 )  & 9.2  & 0.33  & -0.12  &  \\
9 & 64.871719  & 52.980251  &  18.67 ( 0.02 )  &  16.92 ( 0.02 )  &  15.85 ( 0.01 )  &  14.12 ( 0.07 )  &  13.93 ( 0.08 )  &  ... (...)  &  ... (...)  & 8.0  & -0.26  & -0.97  &  CL\\
10 & 64.874032  & 52.970242  &  18.68 ( 0.02 )  &  16.97 ( 0.02 )  &  16.03 ( 0.01 )  &  ... (...)  &  14.08 ( 0.08 )  &  ... (...)  &  ... (...)  & 8.5  & -0.35  & -1.08  &  CL\\
11 & 64.874515  & 52.983613  &  14.70 ( 0.01 )  &  14.33 ( 0.01 )  &  14.17 ( 0.01 )  &  13.79 ( 0.05 )  &  13.66 ( 0.08 )  &  ... (...)  &  ... (...)  & 0.0  & -2.13  & -2.13  &  (CL)\\
12 & 64.874823  & 52.963171  &  15.34 ( 0.01 )  &  14.80 ( 0.01 )  &  14.65 ( 0.01 )  &  14.53 ( 0.05 )  &  14.62 ( 0.09 )  &  ... (...)  &  ... (...)  & 0.0  & -2.71  & -2.71  &  \\
13 & 64.875201  & 52.986769  &  14.58 ( 0.01 )  &  14.12 ( 0.01 )  &  13.97 ( 0.01 )  &  13.76 ( 0.08 )  &  13.72 ( 0.09 )  &  ... (...)  &  ... (...)  & 0.0  & -2.46  & -2.46  &  (CL)\\
14 & 64.875238  & 52.976927  &  14.18 ( 0.01 )  &  13.92 ( 0.01 )  &  13.78 ( 0.01 )  &  13.51 ( 0.06 )  &  13.54 ( 0.06 )  &  ... (...)  &  ... (...)  & 0.0  & -2.44  & -2.44  &  (CL)\\
15 & 64.876513  & 52.979908  &  16.34 ( 0.01 )  &  15.00 ( 0.01 )  &  14.21 ( 0.01 )  &  12.99 ( 0.05 )  &  12.27 ( 0.04 )  &  ... (...)  &  ... (...)  & 4.7  & -0.38  & -0.80  &  CL\\
16 & 64.877465  & 52.996965  &  18.14 ( 0.02 )  &  16.82 ( 0.01 )  &  16.06 ( 0.01 )  &  15.38 ( 0.10 )  &  ... (...)  &  ... (...)  &  ... (...)  & 4.8  & -1.48  & -1.96  &  \\
17 & 64.877614  & 52.968560  &  17.25 ( 0.01 )  &  15.41 ( 0.01 )  &  14.06 ( 0.01 )  &  12.45 ( 0.07 )  &  11.54 ( 0.03 )  &  ... (...)  &  ... (...)  & 7.1  & 0.34  & -0.29  &  CL\\
18 & 64.877991  & 52.984658  &  17.94 ( 0.01 )  &  16.37 ( 0.01 )  &  15.40 ( 0.01 )  &  14.25 ( 0.07 )  &  13.69 ( 0.06 )  &  ... (...)  &  ... (...)  & 6.5  & -0.64  & -1.21  &  CL\\
19 & 64.878377  & 52.978171  &  17.38 ( 0.01 )  &  15.82 ( 0.01 )  &  14.90 ( 0.01 )  &  ... (...)  &  13.28 ( 0.06 )  &  ... (...)  &  ... (...)  & 6.7  & -0.77  & -1.35  &  CL\\
20 & 64.878445  & 52.978744  &  18.94 ( 0.04 )  &  16.51 ( 0.02 )  &  15.02 ( 0.01 )  &  ... (...)  &  12.77 ( 0.05 )  &  ... (...)  &  ... (...)  & 13.7  & 0.02  & -1.16  &  CL\\
21 & 64.878657  & 52.976391  &  17.19 ( 0.01 )  &  15.92 ( 0.01 )  &  15.15 ( 0.01 )  &  13.80 ( 0.09 )  &  13.44 ( 0.07 )  &  ... (...)  &  ... (...)  & 4.0  & -0.59  & -0.94  &  CL\\
22 & 64.879152  & 53.010537  &  16.07 ( 0.01 )  &  15.64 ( 0.01 )  &  15.48 ( 0.01 )  &  15.49 ( 0.10 )  &  ... (...)  &  ... (...)  &  ... (...)  & 0.0  & -2.75  & -2.75  &  \\
23 & 64.880502  & 52.981004  &  18.22 ( 0.02 )  &  16.93 ( 0.01 )  &  16.15 ( 0.01 )  &  ... (...)  &  14.53 ( 0.09 )  &  ... (...)  &  ... (...)  & 4.2  & -0.77  & -1.13  &  CL\\
24 & 64.882245  & 52.984782  &  18.56 ( 0.03 )  &  17.19 ( 0.02 )  &  16.43 ( 0.01 )  &  ... (...)  &  14.59 ( 0.09 )  &  ... (...)  &  ... (...)  & 5.3  & -0.49  & -0.94  &  CL\\
25 & 64.882347  & 52.974461  &  17.59 ( 0.02 )  &  17.00 ( 0.03 )  &  16.40 ( 0.06 )  &  13.19 ( 0.10 )  &  ... (...)  &  ... (...)  &  ... (...)  & 0.0  & 3.15  & 3.15  &  CL\\
26 & 64.882762 & 52.976554 & 20.43 ( 0.17 ) & 17.98 ( 0.07 ) & 16.15 ( 0.04 ) & 12.14 ( 0.06 ) & 10.33 ( 0.05 ) & 8.40 ( 0.04 ) & ... (...) & 11.4 & 4.48 & 3.69 & CL\\
27 & 64.882969  & 53.010907  &  15.63 ( 0.01 )  &  15.17 ( 0.01 )  &  15.03 ( 0.01 )  &  14.99 ( 0.09 )  &  14.86 ( 0.09 )  &  ... (...)  &  ... (...)  & 0.0  & -2.59  & -2.59  &  \\
28 & 64.883575  & 52.981405  &  16.89 ( 0.01 )  &  15.46 ( 0.01 )  &  14.76 ( 0.01 )  &  13.90 ( 0.05 )  &  13.81 ( 0.07 )  &  ... (...)  &  ... (...)  & 6.8  & -1.53  & -2.12  &  CL\\
29 & 64.883634  & 52.977213  &  17.72 ( 0.02 )  &  16.07 ( 0.01 )  &  15.06 ( 0.02 )  &  13.55 ( 0.05 )  &  12.68 ( 0.06 )  &  ... (...)  &  ... (...)  & 7.1  & 0.15  & -0.47  &  CL\\
30 & 64.884251  & 52.955171  &  16.16 ( 0.01 )  &  15.59 ( 0.01 )  &  15.36 ( 0.01 )  &  15.33 ( 0.09 )  &  ... (...)  &  ... (...)  &  ... (...)  & 0.0  & -2.68  & -2.68  &  \\
31 & 64.884275  & 52.994509  &  15.18 ( 0.01 )  &  14.72 ( 0.01 )  &  14.50 ( 0.01 )  &  14.26 ( 0.07 )  &  14.43 ( 0.07 )  &  ... (...)  &  ... (...)  & 0.0  & -2.64  & -2.64  &  \\
32 & 64.884537  & 52.976381  &  17.60 ( 0.01 )  &  15.52 ( 0.01 )  &  14.29 ( 0.01 )  &  12.75 ( 0.05 )  &  12.08 ( 0.04 )  &  ... (...)  &  ... (...)  & 11.2  & -0.01  & -1.00  &  CL\\
33 & 64.884786  & 52.981426  &  17.24 ( 0.01 )  &  16.50 ( 0.01 )  &  16.04 ( 0.02 )  &  14.45 ( 0.08 )  &  13.79 ( 0.07 )  &  ... (...)  &  ... (...)  & 0.0  & 0.05  & 0.05  &  CL\\
34 & 64.885295  & 52.978710  &  13.71 ( 0.01 )  &  13.06 ( 0.01 )  &  12.60 ( 0.01 )  &  11.40 ( 0.06 )  &  10.95 ( 0.06 )  &  9.51 ( 0.06 )  &  ... (...)  & 0.0  & -0.08  & -0.08  &  CL\\
35 & 64.885327  & 52.955001  &  15.71 ( 0.01 )  &  15.05 ( 0.01 )  &  14.89 ( 0.01 )  &  14.90 ( 0.07 )  &  14.88 ( 0.09 )  &  ... (...)  &  ... (...)  & 0.7  & -2.77  & -2.83  &  \\
36 & 64.885802  & 52.978572  &  14.12 ( 0.01 )  &  13.89 ( 0.01 )  &  13.13 ( 0.01 )  &  ... (...)  &  ... (...)  &  9.48 ( 0.05 )  &  ... (...)  & 0.0  & 0.64  & 0.64  &  CL\\
37 & 64.885903  & 52.981400  &  18.59 ( 0.03 )  &  17.64 ( 0.06 )  &  16.41 ( 0.02 )  &  14.62 ( 0.09 )  &  14.25 ( 0.09 )  &  ... (...)  &  ... (...)  & 0.0  & 0.01  & 0.01  &  CL\\
38 & 64.886043  & 52.983772  &  17.19 ( 0.01 )  &  15.96 ( 0.01 )  &  14.92 ( 0.01 )  &  13.52 ( 0.06 )  &  12.89 ( 0.06 )  &  ... (...)  &  ... (...)  & 1.5  & -0.23  & -0.36  &  CL\\
39 & 64.886235  & 52.975407  &  16.03 ( 0.01 )  &  14.68 ( 0.01 )  &  13.94 ( 0.01 )  &  12.85 ( 0.04 )  &  12.26 ( 0.06 )  &  ... (...)  &  ... (...)  & 5.1  & -0.70  & -1.15  &  CL\\
40 & 64.886244  & 52.976270  &  18.37 ( 0.03 )  &  16.83 ( 0.03 )  &  15.93 ( 0.06 )  &  ... (...)  &  13.55 ( 0.10 )  &  ... (...)  &  ... (...)  & 6.5  & 0.18  & -0.38  &  CL\\
41 & 64.886631  & 52.976484  &  18.28 ( 0.02 )  &  17.03 ( 0.04 )  &  16.04 ( 0.04 )  &  ... (...)  &  13.55 ( 0.10 )  &  ... (...)  &  ... (...)  & 2.0  & 0.33  & 0.16  &  CL\\
42 & 64.886719  & 52.970848  &  18.30 ( 0.02 )  &  16.34 ( 0.01 )  &  15.33 ( 0.02 )  &  ... (...)  &  13.59 ( 0.09 )  &  ... (...)  &  ... (...)  & 11.3  & -0.61  & -1.59  &  CL\\
43 & 64.887113  & 52.970847  &  18.71 ( 0.03 )  &  16.84 ( 0.02 )  &  15.80 ( 0.01 )  &  ... (...)  &  13.59 ( 0.09 )  &  ... (...)  &  ... (...)  & 9.9  & -0.02  & -0.88  &  CL\\
44 & 64.887157  & 52.978165  &  11.90 ( 0.01 )  &  ... (...)  &  11.86 ( 0.01 )  &  11.29 ( 0.04 )  &  10.97 ( 0.04 )  &  9.79 ( 0.09 )  &  ... (...)  & … & -1.02  & … &  (CL)\\
45 & 64.887611  & 52.979913  &  15.75 ( 0.01 )  &  14.61 ( 0.01 )  &  14.06 ( 0.01 )  &  13.27 ( 0.07 )  &  12.62 ( 0.06 )  &  ... (...)  &  ... (...)  & 3.8  & -1.04  & -1.38  &  CL\\
46 & 64.887934  & 52.976772  &  16.62 ( 0.01 )  &  15.36 ( 0.01 )  &  14.62 ( 0.01 )  &  ... (...)  &  13.23 ( 0.07 )  &  ... (...)  &  ... (...)  & 4.1  & -1.06  & -1.41  &  CL\\
47 & 64.888214  & 52.982900  &  16.49 ( 0.01 )  &  15.40 ( 0.01 )  &  14.96 ( 0.01 )  &  14.61 ( 0.10 )  &  ... (...)  &  ... (...)  &  ... (...)  & 4.2  & -2.08  & -2.49  &  CL\\
48 & 64.889264  & 53.015272  &  14.49 ( 0.01 )  &  13.96 ( 0.01 )  &  13.81 ( 0.01 )  &  13.76 ( 0.07 )  &  13.63 ( 0.06 )  &  ... (...)  &  ... (...)  & 0.0  & -2.57  & -2.57  &  \\
49 & 64.889533  & 52.974725  &  18.91 ( 0.04 )  &  17.56 ( 0.03 )  &  16.36 ( 0.01 )  &  ... (...)  &  13.93 ( 0.10 )  &  ... (...)  &  ... (...)  & 1.7  & 0.25  & 0.11  &  CL\\
50 & 64.889777  & 52.986988  &  17.14 ( 0.01 )  &  16.05 ( 0.01 )  &  15.50 ( 0.01 )  &  14.72 ( 0.09 )  &  14.18 ( 0.08 )  &  ... (...)  &  ... (...)  & 3.4  & -1.17  & -1.47  &  CL\\
51 & 64.889832  & 52.972521  &  17.35 ( 0.01 )  &  15.45 ( 0.01 )  &  14.25 ( 0.01 )  &  12.43 ( 0.06 )  &  11.81 ( 0.06 )  &  ... (...)  &  ... (...)  & 9.1  & 0.31  & -0.49  &  CL\\
52 & 64.890393  & 52.974944  &  17.42 ( 0.02 )  &  16.12 ( 0.01 )  &  15.45 ( 0.02 )  &  ... (...)  &  13.74 ( 0.07 )  &  ... (...)  &  ... (...)  & 5.0  & -0.65  & -1.08  &  CL\\
53 & 64.890419  & 52.982329  &  17.26 ( 0.01 )  &  16.18 ( 0.02 )  &  15.49 ( 0.02 )  &  ... (...)  &  13.96 ( 0.09 )  &  ... (...)  &  ... (...)  & 2.2  & -0.88  & -1.06  &  CL\\
54 & 64.890541  & 52.988262  &  18.73 ( 0.03 )  &  16.98 ( 0.02 )  &  15.85 ( 0.01 )  &  ... (...)  &  13.58 ( 0.06 )  &  ... (...)  &  ... (...)  & 7.6  & 0.04  & -0.61  &  CL\\
55 & 64.890649  & 52.973240  &  12.29 ( 0.01 )  &  12.66 ( 0.01 )  &  12.20 ( 0.01 )  &  11.59 ( 0.06 )  &  11.45 ( 0.04 )  &  ... (...)  &  ... (...)  & 0.0  & -1.82  & -1.82  &  (CL)\\
56 & 64.890728  & 52.983445  &  16.20 ( 0.01 )  &  14.80 ( 0.01 )  &  13.85 ( 0.01 )  &  12.24 ( 0.05 )  &  11.53 ( 0.03 )  &  10.24 ( 0.08 )  &  ... (...)  & 4.4  & 0.49  & 0.19  &  CL\\
57 & 64.891366  & 52.988806  &  17.62 ( 0.01 )  &  15.80 ( 0.01 )  &  14.59 ( 0.01 )  &  13.05 ( 0.04 )  &  12.56 ( 0.04 )  &  ... (...)  &  ... (...)  & 7.8  & -0.20  & -0.89  &  CL\\
58 & 64.892107  & 52.981927  &  14.85 ( 0.01 )  &  14.36 ( 0.01 )  &  13.96 ( 0.01 )  &  13.53 ( 0.09 )  &  13.25 ( 0.08 )  &  ... (...)  &  ... (...)  & 0.0  & -1.90  & -1.90  &  (CL)\\
59 & 64.892278  & 52.967855  &  ... (...)  &  ... (...)  &  ... (...)  &  ... (...)  &  12.73 ( 0.10 )  &  10.55 ( 0.09 )  &  ... (...)  &  ...  & 5.40  & … &  CL\\
60 & 64.892469  & 52.989916  &  18.49 ( 0.02 )  &  16.89 ( 0.01 )  &  15.86 ( 0.01 )  &  ... (...)  &  14.23 ( 0.08 )  &  ... (...)  &  ... (...)  & 6.2  & -0.74  & -1.28  &  CL\\
61 & 64.892779  & 52.988178  &  18.65 ( 0.02 )  &  16.94 ( 0.01 )  &  15.78 ( 0.01 )  &  14.06 ( 0.09 )  &  13.43 ( 0.08 )  &  ... (...)  &  ... (...)  & 6.9  & 0.19  & -0.42  &  CL\\
62 & 64.894525  & 52.965333  &  18.64 ( 0.02 )  &  15.85 ( 0.01 )  &  14.00 ( 0.01 )  &  11.79 ( 0.04 )  &  10.97 ( 0.04 )  &  9.40 ( 0.04 )  &  9.66 ( 0.05 )  & 15.7  & 0.53  & -0.24  &  CL\\
63 & 64.894563  & 52.960684  &  16.82 ( 0.01 )  &  16.07 ( 0.01 )  &  15.62 ( 0.01 )  &  15.36 ( 0.10 )  &  ... (...)  &  ... (...)  &  ... (...)  & 0.0  & -2.25  & -2.25  &  \\
64 & 64.895302  & 52.977628  &  17.37 ( 0.01 )  &  16.34 ( 0.01 )  &  15.60 ( 0.01 )  &  14.50 ( 0.07 )  &  13.72 ( 0.06 )  &  ... (...)  &  ... (...)  & 1.2  & -0.49  & -0.59  &  CL\\
65 & 64.896202  & 52.985814  &  14.38 ( 0.01 )  &  13.93 ( 0.01 )  &  13.74 ( 0.01 )  &  13.40 ( 0.05 )  &  13.28 ( 0.07 )  &  ... (...)  &  ... (...)  & 0.0  & -2.20  & -2.20  &  (CL)\\
66 & 64.897239  & 52.987433  &  15.98 ( 0.01 )  &  15.25 ( 0.01 )  &  15.01 ( 0.01 )  &  14.55 ( 0.09 )  &  ... (...)  &  ... (...)  &  ... (...)  & 0.8  & -1.88  & -1.96  &  (CL)\\
67 & 64.897705  & 53.008391  &  16.40 ( 0.01 )  &  15.63 ( 0.01 )  &  15.28 ( 0.01 )  &  15.15 ( 0.09 )  &  ... (...)  &  ... (...)  &  ... (...)  & 0.6  & -2.49  & -2.55  &  \\
68 & 64.898921  & 53.012716  &  15.78 ( 0.01 )  &  15.24 ( 0.01 )  &  15.06 ( 0.01 )  &  14.96 ( 0.08 )  &  ... (...)  &  ... (...)  &  ... (...)  & 0.0  & -2.55  & -2.55  &  \\
69 & 64.898923  & 53.012726  &  15.78 ( 0.01 )  &  15.24 ( 0.01 )  &  15.06 ( 0.01 )  &  14.96 ( 0.08 )  &  ... (...)  &  ... (...)  &  ... (...)  & 0.0  & -2.56  & -2.56  &  \\
70 & 64.900747  & 53.011083  &  15.73 ( 0.01 )  &  14.97 ( 0.01 )  &  14.79 ( 0.01 )  &  14.82 ( 0.09 )  &  14.63 ( 0.10 )  &  ... (...)  &  ... (...)  & 1.8  & -2.63  & -2.78  &  \\
71 & 64.900758  & 53.011094  &  15.72 ( 0.01 )  &  14.97 ( 0.01 )  &  14.79 ( 0.01 )  &  14.82 ( 0.09 )  &  14.63 ( 0.10 )  &  ... (...)  &  ... (...)  & 1.6  & -2.63  & -2.77  &  \\
72 & 64.902540  & 52.948835  &  14.09 ( 0.01 )  &  13.38 ( 0.01 )  &  13.19 ( 0.01 )  &  13.23 ( 0.05 )  &  13.23 ( 0.05 )  &  ... (...)  &  ... (...)  & 1.1  & -2.84  & -2.93  &  \\
73 & 64.903706  & 52.996184  &  14.74 ( 0.01 )  &  14.13 ( 0.01 )  &  13.95 ( 0.01 )  &  13.84 ( 0.08 )  &  13.79 ( 0.06 )  &  ... (...)  &  ... (...)  & 0.0  & -2.57  & -2.57  &  \\
74 & 64.906434  & 52.951374  &  17.04 ( 0.01 )  &  16.33 ( 0.01 )  &  16.02 ( 0.01 )  &  15.68 ( 0.09 )  &  ... (...)  &  ... (...)  &  ... (...)  & 0.1  & -2.12  & -2.13  &  \\
75 & 64.907308  & 52.952973  &  16.17 ( 0.01 )  &  15.43 ( 0.01 )  &  15.05 ( 0.01 )  &  15.02 ( 0.06 )  &  14.73 ( 0.09 )  &  ... (...)  &  ... (...)  & 0.0  & -2.42  & -2.42  &  \\
76 & 64.908143  & 52.983768  &  15.07 ( 0.01 )  &  14.52 ( 0.01 )  &  14.31 ( 0.01 )  &  13.96 ( 0.05 )  &  13.99 ( 0.07 )  &  ... (...)  &  ... (...)  & 0.0  & -2.35  & -2.35  &  \\
77 & 64.910304  & 52.958083  &  16.35 ( 0.01 )  &  15.83 ( 0.01 )  &  15.67 ( 0.01 )  &  15.42 ( 0.08 )  &  ... (...)  &  ... (...)  &  ... (...)  & 0.0  & -2.29  & -2.29  &  \\
78 & 64.911027  & 52.960540  &  15.14 ( 0.01 )  &  14.45 ( 0.01 )  &  13.94 ( 0.01 )  &  13.21 ( 0.04 )  &  13.00 ( 0.04 )  &  ... (...)  &  ... (...)  & 0.0  & -1.58  & -1.58  &  \\
79 & 64.912288  & 52.998403  &  17.14 ( 0.01 )  &  16.10 ( 0.01 )  &  15.67 ( 0.01 )  &  15.43 ( 0.09 )  &  ... (...)  &  ... (...)  &  ... (...)  & 3.7  & -2.29  & -2.65  &  \\
80 & 64.916342  & 52.958633  &  16.28 ( 0.01 )  &  15.69 ( 0.01 )  &  15.41 ( 0.01 )  &  15.23 ( 0.10 )  &  ... (...)  &  ... (...)  &  ... (...)  & 0.0  & -2.41  & -2.41  &  \\
81 & 64.916802  & 52.973346  &  14.72 ( 0.01 )  &  14.34 ( 0.01 )  &  14.20 ( 0.01 )  &  14.20 ( 0.05 )  &  14.15 ( 0.08 )  &  ... (...)  &  ... (...)  & 0.0  & -2.73  & -2.73  &  \\
82 & 64.916847  & 53.001318  &  13.62 ( 0.01 )  &  13.10 ( 0.01 )  &  13.26 ( 0.01 )  &  12.96 ( 0.04 )  &  12.93 ( 0.04 )  &  ... (...)  &  ... (...)  & 1.3  & -2.35  & -2.46  &  \\
83 & 64.917298  & 52.982781  &  17.09 ( 0.01 )  &  16.25 ( 0.01 )  &  15.91 ( 0.01 )  &  15.01 ( 0.10 )  &  ... (...)  &  ... (...)  &  ... (...)  & 1.6  & -1.07  & -1.23  &  \\
84 & 64.918288  & 52.995132  &  14.71 ( 0.01 )  &  13.71 ( 0.01 )  &  13.26 ( 0.01 )  &  12.98 ( 0.03 )  &  12.94 ( 0.04 )  &  ... (...)  &  ... (...)  & 3.0  & -2.36  & -2.62  &  \\
85 & 64.919530  & 52.971248  &  15.08 ( 0.01 )  &  14.63 ( 0.01 )  &  14.45 ( 0.01 )  &  14.49 ( 0.06 )  &  14.45 ( 0.09 )  &  ... (...)  &  ... (...)  & 0.0  & -2.79  & -2.79  &  \\
86 & 64.924564  & 52.995167  &  13.56 ( 0.01 )  &  13.10 ( 0.01 )  &  13.13 ( 0.01 )  &  12.59 ( 0.03 )  &  12.53 ( 0.03 )  &  ... (...)  &  ... (...)  & 0.0  & -1.99  & -1.99  &  \\
87 & 64.925814  & 52.960782  &  16.79 ( 0.01 )  &  16.13 ( 0.01 )  &  15.82 ( 0.01 )  &  15.80 ( 0.10 )  &  ... (...)  &  ... (...)  &  ... (...)  & 0.0  & -2.71  & -2.71  &  \\
88 & 64.926157  & 52.963839  &  15.07 ( 0.01 )  &  14.73 ( 0.01 )  &  14.57 ( 0.01 )  &  14.62 ( 0.08 )  &  14.47 ( 0.06 )  &  ... (...)  &  ... (...)  & 0.0  & -2.68  & -2.68  &  \\
89 & 64.926949  & 52.973223  &  14.82 ( 0.01 )  &  14.51 ( 0.01 )  &  14.39 ( 0.01 )  &  14.39 ( 0.05 )  &  14.38 ( 0.10 )  &  ... (...)  &  ... (...)  & 0.0  & -2.76  & -2.76  &  \\
90 & 64.927552  & 52.956403  &  15.88 ( 0.01 )  &  15.35 ( 0.01 )  &  15.22 ( 0.01 )  &  15.23 ( 0.08 )  &  ... (...)  &  ... (...)  &  ... (...)  & 0.0  & -2.76  & -2.76  &  \\
91 & 64.929206  & 52.983985  &  16.60 ( 0.01 )  &  15.89 ( 0.01 )  &  15.59 ( 0.01 )  &  15.22 ( 0.08 )  &  ... (...)  &  ... (...)  &  ... (...)  & 0.3  & -2.06  & -2.08  &  \\
92 & 64.932245  & 53.000599  &  12.99 ( 0.01 )  &  12.82 ( 0.01 )  &  12.83 ( 0.01 )  &  12.55 ( 0.04 )  &  12.61 ( 0.04 )  &  ... (...)  &  ... (...)  & 0.0  & -2.47  & -2.47  &  \\
93 & 64.937640  & 52.990471  &  14.84 ( 0.01 )  &  14.65 ( 0.01 )  &  14.51 ( 0.01 )  &  14.40 ( 0.07 )  &  14.53 ( 0.07 )  &  ... (...)  &  ... (...)  & 0.0  & -2.76  & -2.76  &  \\
94 & 64.937836  & 52.978751  &  16.17 ( 0.01 )  &  15.42 ( 0.01 )  &  15.19 ( 0.01 )  &  15.08 ( 0.07 )  &  15.16 ( 0.10 )  &  ... (...)  &  ... (...)  & 1.3  & -2.72  & -2.83  &  \\
95 & 64.946757  & 52.985985  &  16.32 ( 0.01 )  &  15.55 ( 0.01 )  &  15.30 ( 0.01 )  &  15.21 ( 0.09 )  &  15.09 ( 0.08 )  &  ... (...)  &  ... (...)  & 1.3  & -2.53  & -2.64  &  \\
96 & 64.946790  & 52.985996  &  16.31 ( 0.01 )  &  15.55 ( 0.01 )  &  15.30 ( 0.01 )  &  15.21 ( 0.09 )  &  15.09 ( 0.08 )  &  ... (...)  &  ... (...)  & 1.2  & -2.52  & -2.63  &  \\
 \enddata
\tablecomments{Column (1): ID number. Columns (2) and (3): right
ascension (RA) and declination (Dec).
Columns (4)--(6): NIR {magnitudes.}
{\it J} band in Column (4), {\it H}
band in Column (5), and $K_S$ band in Column (6).
Magnitude errors are shown in parentheses.
Columns (7)--(10): {\it Spitzer}/IRAC magnitudes. 3.6-$\mu$m band in
Column (7), 4.5-$\mu$m band in Column (8), 5.6-$\mu$m band in Column
(9), and 8.0-$\mu$m band magnitudes in Column (10). Magnitude errors are
shown in parentheses.
Column (11): $A_V$ values are only provided for sources with valid {\it
JHK}$_S$ photometry.
Columns (12)--(13): SED slopes [$d \log (\lambda F_\lambda) / d \log
(\lambda)$] derived from the NIR $K_S$ and IRAC bands.
The SED slopes derived from the observed data for each source ($\alpha$)
are shown in Column (12), and the SED slopes derived from the data
corrected for extinction using the $A_V$ derived for each object
($\alpha_0$) are shown in Column (13).
Column (14): probable cluster members are labeled ``CL,'' whereas
 sources inside the cluster region, which are not identified as cluster
 members are labeled ``(CL).''} 
 
\end{deluxetable*}
\end{longrotatetable}


\clearpage

\end{document}